\documentclass[12pt,letterpaper]{article}
\pdfoutput=1
\usepackage{jheppub}
\usepackage{amsfonts, amsthm}
\usepackage[english]{babel}
\usepackage[utf8]{inputenc}
\usepackage{slashed}
\usepackage{mathrsfs}
\usepackage{amssymb}
\usepackage{color}
\hypersetup{unicode}

\newcommand{\eq}{\begin{equation}}
\newcommand{\feq}{\end{equation}}
\newcommand{\eqn}{\begin{eqnarray}}
\newcommand{\feqn}{\end{eqnarray}}
\newcommand{\eqNN}{\begin{equation*}}
\newcommand{\feqNN}{\end{equation*}}
\font\mybb=msbm10 at 12pt
\def\bb#1{\hbox{\mybb#1}}

\def\bR {\bb{R}}

\newcommand{\D}{{\rm d}}

\title{Rotating black holes in the FI-gauged $N=2$, $D=4$ 
$\overline{\mathbb{C}\text{P}}^n$ model}

\author{Nicoletta Daniele$^1$,}
\author{Federico Faedo$^{1,2}$,}
\author{Dietmar Klemm$^{1,2}$}
\author{and Pedro F.~Ram\'{\i}rez$^2$}

\affiliation{$^1$ Dipartimento di Fisica, Universit\`a di Milano, \\
Via Celoria 16, I-20133 Milano}
\affiliation{$^2$ INFN, Sezione di Milano, \\
Via Celoria 16, I-20133 Milano}

\emailAdd{nicoletta.daniele@studenti.unimi.it}
\emailAdd{federico.faedo@unimi.it}
\emailAdd{dietmar.klemm@mi.infn.it}
\emailAdd{ramirez.pedro@mi.infn.it}
\preprint{IFUM-1072-FT}

\abstract{We construct supersymmetric black holes with rotation
or NUT charge for the $\overline{\mathbb{C}\text{P}}^n$- and the $\text{t}^3$ model of
$N=2$, $D=4$ $\text{U}(1)$ FI-gauged supergravity. The solutions preserve 2 real supercharges,
which are doubled for their near-horizon geometry. For the $\overline{\mathbb{C}\text{P}}^n$
model we also present a generalization to the nonextremal case, which turns out to be
characterized by a Carter-Pleba\'nski-type metric, and has $n+3$ independent parameters,
corresponding to mass, angular momentum as well as $n+1$ magnetic charges. We discuss the thermodynamics of these solutions, obtain a Christodoulou-Ruffini mass formula, and shew that they
obey a first law of thermodynamics and that the product of horizon areas depends on the angular
momentum and the magnetic charges only.
At least some of the BPS black holes that we obtain may become instrumental
for future microscopic entropy computations involving a supersymmetric index.
}

\keywords{Black Holes, AdS/CFT Correspondence, Classical Theories of Gravity, Supergravity Models}

\begin{document}
\maketitle
\flushbottom


\section{Introduction and summary of results}

Black holes in gauged supergravity theories provide an important testground
to address fundamental questions of gravity, both at the classical and quantum level.
In particular, one may be interested in uniqueness- or no hair theorems, the final state of
black hole evolution, or the problem of black hole microstates.
In gauged supergravity, the solutions often (but not always; this depends essentially on whether the
scalar potential has critical points)
have AdS asymptotics, and one can then try to study these issues guided by the AdS/CFT
correspondence. A nice example for this is the recent microscopic entropy
calculation \cite{Benini:2015eyy,Benini:2016hjo,Benini:2016rke,Cabo-Bizet:2017jsl}
for the black hole solutions to $N=2$, $D=4$ Fayet-Iliopoulos (FI)-gauged
supergravity constructed in \cite{Cacciatori:2009iz}. These preserve two real supercharges,
and are dual to a topologically twisted ABJM theory, whose partition function can be computed
exactly using supersymmetric localization techniques. This partition function can also be interpreted as
the Witten index of the superconformal quantum mechanics resulting from dimensionally reducing the
ABJM theory on a Riemann surface. The results of
\cite{Benini:2015eyy,Benini:2016hjo,Benini:2016rke,Cabo-Bizet:2017jsl} represent
the first exact black hole microstate counting that uses AdS/CFT and that does
not involve an $\text{AdS}_3$ factor\footnote{Or geometries related to $\text{AdS}_3$, like those
appearing in the Kerr/CFT correspondence \cite{Guica:2008mu}.} with a corresponding two-dimensional
CFT, whose asymptotic level density is evaluated with the Cardy formula.
Subsequently, this matching was extended to many other examples and in various directions,
see e.~g.~\cite{Hosseini:2016tor,Azzurli:2017kxo,Liu:2017vll,Hosseini:2017fjo,Benini:2017oxt,
Liu:2018bac,Hristov:2018lod} and references therein.

On the other hand, black hole solutions to gauged supergravity are also relevant for a number of recent 
developments in high energy- and in condensed matter physics, since they provide the
dual description of the quark-gluon plasma \cite{Gubser:2009fc} as well as of certain condensed matter
systems at finite temperature (cf.~\cite{Hartnoll:2009sz} for a review) and quantum phase
transitions \cite{Hartnoll:2007ih}. Of particular importance in this context are models that contain
Einstein gravity coupled to $\text{U}(1)$ gauge fields and neutral scalars, which have been
instrumental to study transitions from Fermi-liquid to non-Fermi-liquid behaviour, 
cf.~\cite{Charmousis:2010zz,Iizuka:2011hg} and references therein. Notice that the
necessity of a bulk $\text{U}(1)$ gauge field arises, because a basic ingredient of
realistic condensed matter systems is the presence of a finite density of charge carriers.
Such models are provided by matter-coupled gauged supergravity. Especially we shall be interested
in the $N=2$ $\text{U}(1)$ FI-gauged theory in four dimensions, which contains indeed neutral scalars
as well as abelian gauge fields.

There are thus a number of reasons to extend the spectrum of known black hole solutions to gauged
supergravity. Since there exists by now a rather large amount of literature on this subject, in the
following we will give an overview on existing solutions, which may be useful for the reader in its own right.
In order to avoid escalation we shall thereby restrict our attention to the $N=2$, $D=4$ $\text{U}(1)$
FI-gauged theory only.
To the best of our knowledge, the first paper on this subject was \cite{Duff:1999gh}, where nonextremal
black holes in the stu model were constructed. These carry four charges, which are either all electric
or magnetic. Ref.~\cite{Sabra:1999ux} derives electrically charged 1/2 BPS solutions for arbitrary
prepotential, which unfortunately are naked singularities as soon as the gauge coupling constant is 
nonvanishing. In \cite{Cacciatori:2009iz} (using the classification scheme of \cite{Cacciatori:2008ek}),
the first examples of genuine supersymmetric black holes in AdS$_4$ with nonconstant scalar fields
were presented for the t$^3$ and the stu model. Typically these are magnetically charged and
represent also the prime instance of static BPS black holes in AdS$_4$ with spherical symmetry.
\cite{DallAgata:2010ejj,Hristov:2010ri} elaborate further on the solutions of \cite{Cacciatori:2009iz},
while \cite{Gnecchi:2013mta,Katmadas:2014faa,Halmagyi:2014qza,Klemm:2015xda} and
\cite{Klemm:2012yg,Toldo:2012ec,Klemm:2012vm,Gnecchi:2012kb} generalize them
to other prepotentials (with dyonic gaugings) and finite temperature respectively.
Rotation was added in \cite{Klemm:2011xw} (BPS case, $-iX^0X^1$-model, only magnetic charges),
\cite{Chow:2013gba,Gnecchi:2013mja} (same model, but nonextremal and dyonic) and very recently
in \cite{Hristov:2018spe} (BPS, cubic prepotential and dyonic gauging). NUT-charged supersymmetric
black holes were constructed in \cite{Colleoni:2012jq} for the $-iX^0X^1$-model and in
\cite{Erbin:2015gha} for a cubic prepotential with dyonic gauging.
It is worth noting that there exists also a strange class of black holes whose horizon is noncompact but
nevertheless has finite area \cite{Gnecchi:2013mja,Klemm:2014rda}. These may provide an
interesting testground to address fundamental questions related to black hole physics or holography. 

Many further hitherto unknown solutions might exist, but are very probably difficult to construct by
trying to solve the coupled Einstein-Maxwell-scalar equations. However, the supersymmetric subclass
of them (if it exists) satisfies first order equations, which should facilitate their discovery and explicit 
construction.

In this paper we shall consider the $\overline{\mathbb{C}\text{P}}^n$- and the $\text{t}^3$ model, 
characterized by a quadratic and cubic prepotential respectively. We start in section \ref{sugra} with a
brief review of $N=2$, $D=4$ FI-gauged supergravity as well as a summary of some results of
\cite{Cacciatori:2008ek,Klemm:2010mc}, where the one quarter and one half supersymmetric
backgrounds of the theory were classified. In section \ref{1/4BPS} we apply the recipe of
\cite{Cacciatori:2008ek} to construct rotating extremal BPS black holes in the 
$\overline{\mathbb{C}\text{P}}^n$ model, which preserve two real supercharges. It is shown that the
latter are doubled for the near-horizon geometry. Moreover, we also obtain BPS black holes with NUT
charge in the same model. The following section is dedicated to the prepotential $F=-(X^1)^3/X^0$,
for which we present first a supersymmetric near-horizon solution, which is subsequently extended to
a full black hole geometry. Finally, \ref{sec:non-extr-CPn} contains a generalization of the solutions
in section \ref{1/4BPS} to the nonextremal case, which turns out to be characterized by a
Carter-Pleba\'nski-type metric, and has $n+3$ independent parameters, corresponding to mass,
angular momentum as well as $n+1$ magnetic charges. We also discuss the thermodynamics
of these solutions, obtain a Christodoulou-Ruffini mass formula, and shew that they obey a first
law of thermodynamics and that the product of horizon areas depends on the angular momentum
and the magnetic charges only.

We believe that at least some of the black holes constructed in this paper may become instrumental
for future microscopic entropy computations involving a supersymmetric index, along the lines
of \cite{Benini:2015eyy,Benini:2016hjo,Benini:2016rke,Cabo-Bizet:2017jsl}.

An appendix contains the equations of motion of the theory under consideration.

\newpage

\section{$N=2$, $D=4$ FI-gauged supergravity}
\label{sugra}

\subsection{The theory and BPS equations}

We consider $N=2$, $D=4$ gauged supergravity coupled to $n$ abelian
vector multiplets \cite{Andrianopoli:1996cm}\footnote{Throughout this paper,
we use the notations and conventions of \cite{Freedman:2012zz}.}.
Apart from the vierbein $e^a_{\mu}$, the bosonic field content includes the
vectors $A^I_{\mu}$ enumerated by $I=0,\ldots,n$, and the complex scalars
$z^{\alpha}$ where $\alpha=1,\ldots,n$. These scalars parametrize
a special K\"ahler manifold, i.~e.~, an $n$-dimensional
Hodge-K\"ahler manifold that is the base of a symplectic bundle, with the
covariantly holomorphic sections
\begin{equation}
{\cal V} = \left(\begin{array}{c} X^I \\ F_I\end{array}\right)\,, \qquad
{\cal D}_{\bar\alpha}{\cal V} = \partial_{\bar\alpha}{\cal V}-\frac 12
(\partial_{\bar\alpha}{\cal K}){\cal V}=0\,, \label{sympl-vec}
\end{equation}
where ${\cal K}$ is the K\"ahler potential and ${\cal D}$ denotes the
K\"ahler-covariant derivative. ${\cal V}$ obeys the symplectic constraint
\begin{equation}
\langle {\cal V},\bar{\cal V}\rangle = X^I\bar F_I-F_I\bar X^I=i\,. \label{sympconst}
\end{equation}
To solve this condition, one defines
\begin{equation}
{\cal V}=e^{{\cal K}(z,\bar z)/2}v(z)\,,
\end{equation}
where $v(z)$ is a holomorphic symplectic vector,
\begin{equation}
v(z) = \left(\begin{array}{c} Z^I(z) \\ \frac{\partial}{\partial Z^I}F(Z)
\end{array}\right)\,.
\end{equation}
F is a homogeneous function of degree two, called the prepotential,
whose existence is assumed to obtain the last expression.
The K\"ahler potential is then
\begin{equation}
e^{-{\cal K}(z,\bar z)} = -i\langle v,\bar v\rangle\,.
\end{equation}
The matrix ${\cal N}_{IJ}$ determining the coupling between the scalars
$z^{\alpha}$ and the vectors $A^I_{\mu}$ is defined by the relations
\begin{equation}\label{defN}
F_I = {\cal N}_{IJ}X^J\,, \qquad {\cal D}_{\bar\alpha}\bar F_I = {\cal N}_{IJ}
{\cal D}_{\bar\alpha}\bar X^J\,.
\end{equation}
The bosonic action reads
\begin{eqnarray}
e^{-1}{\cal L}_{\text{bos}} &=& \frac 12R + \frac 14(\text{Im}\,
{\cal N})_{IJ}F^I_{\mu\nu}F^{J\mu\nu} - \frac 18(\text{Re}\,{\cal N})_{IJ}\,e^{-1}
\epsilon^{\mu\nu\rho\sigma}F^I_{\mu\nu}F^J_{\rho\sigma} \nonumber \\
&& -g_{\alpha\bar\beta}\partial_{\mu}z^{\alpha}\partial^{\mu}\bar z^{\bar\beta}
- V\,, \label{action}
\end{eqnarray}
with the scalar potential
\eq
V = -2g^2\xi_I\xi_J[(\text{Im}\,{\cal N})^{-1|IJ}+8\bar X^IX^J]\,, \label{scal-pot}
\feq
that results from U$(1)$ Fayet-Iliopoulos gauging. Here, $g$ denotes the
gauge coupling and the $\xi_I$ are FI constants. In what follows, we define
$g_I\equiv g\xi_I$.

The most general timelike supersymmetric background of the theory described
above was constructed in \cite{Cacciatori:2008ek}, and is given by
\eq
ds^2 = -4|b|^2(dt+\sigma)^2 + |b|^{-2}(dz^2+e^{2\Phi}dwd\bar w)\ , \label{gen-metr}
\feq
where the complex function $b(z,w,\bar w)$, the real function $\Phi(z,w,\bar w)$
and the one-form $\sigma=\sigma_wdw+\sigma_{\bar w}d\bar w$, together with the
symplectic section \eqref{sympl-vec}\footnote{Note that also $\sigma$ and
$\cal V$ are independent of $t$.} are determined by the equations
\eq
\partial_z\Phi = 2ig_I\left(\frac{{\bar X}^I}b-\frac{X^I}{\bar b}\right)\ ,
\label{dzPhi}
\feq
\begin{eqnarray}
&&\qquad 4\partial\bar\partial\left(\frac{X^I}{\bar b}-\frac{\bar X^I}b\right) + \partial_z\left[e^{2\Phi}\partial_z
\left(\frac{X^I}{\bar b}-\frac{\bar X^I}b\right)\right]  \label{bianchi} \\
&&-2ig_J\partial_z\left\{e^{2\Phi}\left[|b|^{-2}(\text{Im}\,{\cal N})^{-1|IJ}
+ 2\left(\frac{X^I}{\bar b}+\frac{\bar X^I}b\right)\left(\frac{X^J}{\bar b}+\frac{\bar X^J}b\right)\right]\right\}= 0\,,
\nonumber
\end{eqnarray}
\begin{eqnarray}
&&\qquad 4\partial\bar\partial\left(\frac{F_I}{\bar b}-\frac{\bar F_I}b\right) + \partial_z\left[e^{2\Phi}\partial_z
\left(\frac{F_I}{\bar b}-\frac{\bar F_I}b\right)\right] \nonumber \\
&&-2ig_J\partial_z\left\{e^{2\Phi}\left[|b|^{-2}\text{Re}\,{\cal N}_{IL}(\text{Im}\,{\cal N})^{-1|JL}
+ 2\left(\frac{F_I}{\bar b}+\frac{\bar F_I}b\right)\left(\frac{X^J}{\bar b}+\frac{\bar X^J}b\right)\right]\right\}
\nonumber \\
&&-8ig_I e^{2\Phi}\left[\langle {\cal I}\,,\partial_z {\cal I}\rangle-\frac{g_J}{|b|^2}\left(\frac{X^J}{\bar b}
+\frac{\bar X^J}b\right)\right] = 0\,, \label{maxwell}
\end{eqnarray}
\begin{equation}
2\partial\bar\partial\Phi=e^{2\Phi}\left[ig_I\partial_z\left(\frac{X^I}{\bar b}-\frac{\bar X^I}b\right)
+\frac2{|b|^2}g_Ig_J(\text{Im}\,{\cal N})^{-1|IJ}+4\left(\frac{g_I X^I}{\bar b}+\frac{g_I \bar X^I}b
\right)^2\right]\,, \label{Delta-Phi}
\end{equation}
\begin{equation}
d\sigma + 2\,\star^{(3)}\!\langle{\cal I}\,,d{\cal I}\rangle - \frac i{|b|^2}g_I\left(\frac{\bar X^I}b
+\frac{X^I}{\bar b}\right)e^{2\Phi}dw\wedge d\bar w=0\,. \label{dsigma}
\end{equation}
Here $\star^{(3)}$ is the Hodge star on the three-dimensional base with metric\footnote{Whereas
in the ungauged case, this base space is flat and thus has trivial holonomy, here we have U(1)
holonomy with torsion \cite{Cacciatori:2008ek}.}
\eq
ds_3^2 = dz^2+e^{2\Phi}dwd\bar w\,, \label{metr-base}
\feq
and we defined $\partial=\partial_w$, $\bar\partial=\partial_{\bar w}$, as well as
\begin{equation}
{\cal I} = \text{Im}\left({\cal V}/\bar b\right)\,, \qquad {\cal R} = \text{Re}\left({\cal V}/\bar b\right)\,.
\end{equation}
Note that the eqns.~\eqref{dzPhi}-\eqref{Delta-Phi} can be written compactly in the symplectically
covariant form
\eq
\partial_z\Phi = 4\langle{\cal I},{\cal G}\rangle\,, \label{partialzPhi}
\feq
\begin{eqnarray}
\Delta{\cal I} &+& 2 e^{-2\Phi}\partial_z\left\{e^{2\Phi}\left[\langle{\cal R},{\cal I}\rangle\Omega{\cal M}
{\cal G} - 4{\cal R}\langle{\cal R},{\cal G}\rangle\right]\right\} \nonumber \\
&-& 4{\cal G}\left[\langle{\cal I},\partial_z
{\cal I}\rangle + 4\langle{\cal R},{\cal I}\rangle\langle{\cal R},{\cal G}\rangle\right] = 0\,, \label{Delta-I}
\end{eqnarray}
\eq
\Delta\Phi = -8\langle{\cal R},{\cal I}\rangle\left[{\cal G}^t{\cal M}{\cal G} + 8|{\cal L}|^2\right]
= 4\langle{\cal R},{\cal I}\rangle V\,, \label{Delta-Phi-cov}
\feq
where ${\cal G}=(g^I,g_I)^t$ represents the symplectic vector of gauge couplings\footnote{In the case
considered here with electric gaugings only, one has $g^I=0$.}, ${\cal L}=\langle{\cal V},{\cal G}\rangle$,
$\Delta$ denotes the covariant Laplacian associated to the base space metric \eqref{metr-base}, and $V$
in \eqref{Delta-Phi-cov} is the scalar potential \eqref{scal-pot}. Moreover,
\eq
\Omega = \left(\begin{array}{cc} 0 & 1 \\ -1 & 0\end{array}\right)\,, \qquad {\cal M} =
\left(\begin{array}{cc}\text{Im}\,{\cal N} + \text{Re}\,{\cal N}(\text{Im}\,{\cal N})^{-1}\text{Re}\,{\cal N} & 
-\text{Re}\,{\cal N}(\text{Im}\,{\cal N})^{-1} \\ -(\text{Im}\,{\cal N})^{-1}\text{Re}\,{\cal N} &
(\text{Im}\,{\cal N})^{-1}\end{array}\right)\,.
\feq
Finally, \eqref{dsigma} can be rewritten as
\eq
d\sigma + \star_h\left(d\Sigma - A + \frac12\nu\Sigma\right) = 0\,, \label{gen-mon}
\feq
where the function $\Sigma$ and the one-form $\nu$ are respectively given by
\eq
\Sigma = \frac i2\ln\frac{\bar b}b\,, \qquad \nu = \frac8\Sigma\langle{\cal G},{\cal R}\rangle dz\,,
\feq
$A$ is the gauge field of the K\"ahler U$(1)$,
\eq
A_{\mu} = -\frac i2(\partial_{\alpha}{\cal K}\partial_{\mu}z^{\alpha} -
         \partial_{\bar\alpha}{\cal K}\partial_{\mu}{\bar z}^{\bar\alpha})\,,
\feq
and $\star_h$ denotes the Hodge star on the Weyl-rescaled base space metric
\eq
h_{ij}dx^i dx^j = \frac1{|b|^4}(dz^2+e^{2\Phi}dwd\bar w)\,.
\feq
\eqref{gen-mon} is the generalized monopole equation \cite{Jones:1985pla}, or more precisely
a K\"ahler-covariant generalization thereof, due to the presence of the one-form $A$.
In order to cast \eqref{dsigma} into the form \eqref{gen-mon}, one has to use the special K\"ahler
identities
\eq\label{identitiesDVV}
\langle{\cal D}_\alpha{\cal V},{\cal V}\rangle = \langle{\cal D}_\alpha{\cal V},\bar{\cal V}\rangle = 0\,.
\feq
Note that\eqref{gen-mon} is invariant under Weyl rescaling, accompanied by a gauge transformation
of $\nu$,
\begin{align}
h_{mn}\D x^m\D x^n \mapsto e^{2\psi}h_{mn}\D x^m\D x^n\,, \quad \Sigma \mapsto
e^{-\psi}\Sigma\,, \quad \nu \mapsto \nu + 2\D\psi\,, \quad A \mapsto e^{-\psi}A\,.
\end{align}
It would be very interesting to better understand the deeper origin of the conformal invariance
of \eqref{gen-mon} in the present context.

The integrability condition for \eqref{gen-mon} reads
\eq
D_i[h^{ij}\sqrt h (D_j - A_j)\Sigma] = 0\,, \label{int-cond}
\feq
with the Weyl-covariant derivative
\eq
D_i = \partial_i - \frac m2\nu_i\,,
\feq
where $m$ denotes the Weyl weight of the corresponding field\footnote{A field $\Gamma$ with Weyl
weight $m$ transforms as $\Gamma\mapsto e^{m\psi}\Gamma$ under a Weyl rescaling.}.
It is straightforward to show that \eqref{int-cond} is equivalent to
\eq
\langle{\cal I},\Delta{\cal I}\rangle + 4 e^{-2\Phi}\partial_z\left(e^{2\Phi}\langle{\cal I},{\cal R}\rangle
\langle{\cal G},{\cal R}\rangle\right) = 0\,,
\feq
which follows from \eqref{Delta-I} by taking the symplectic product with $\cal I$.
To shew this, one has to use
\eq
\frac12\left({\cal M} + i\Omega\right) = \Omega\bar{\cal V}{\cal V}\Omega +
\Omega {\cal D}_\alpha {\cal V} g^{\alpha\bar\beta} {\cal D}_{\bar\beta}\bar{\cal V}\Omega,
\feq
\eq
\langle {\cal D}_\alpha {\cal V}, {\cal D}_\beta {\cal V}\rangle = 0\,, \qquad
\langle {\cal D}_\alpha {\cal V}, {\cal D}_{\bar\beta}\bar{\cal V}\rangle = -i g_{\alpha\bar\beta}\,,
\feq
as well as
\eqref{partialzPhi} and \eqref{identitiesDVV}.

Given $b$, $\Phi$, $\sigma$ and $\cal V$, the fluxes read
\begin{eqnarray}
F^I&=&2(dt+\sigma)\wedge d\left[bX^I+\bar b\bar X^I\right]+|b|^{-2}dz\wedge d\bar w
\left[\bar X^I(\bar\partial\bar b+iA_{\bar w}\bar b)+({\cal D}_{\alpha}X^I)b\bar\partial z^{\alpha}-
\right. \nonumber \\
&&\left. X^I(\bar\partial b-iA_{\bar w}b)-({\cal D}_{\bar\alpha}\bar X^I)\bar b\bar\partial\bar z^{\bar\alpha}
\right]-|b|^{-2}dz\wedge dw\left[\bar X^I(\partial\bar b+iA_w\bar b)+\right. \nonumber \\
&&\left.({\cal D}_{\alpha}X^I)b\partial z^{\alpha}-X^I(\partial b-iA_w b)-({\cal D}_{\bar\alpha}\bar X^I)
\bar b\partial\bar z^{\bar\alpha}\right]- \nonumber \\
&&\frac 12|b|^{-2}e^{2\Phi}dw\wedge d\bar w\left[\bar X^I(\partial_z\bar b+iA_z\bar b)+({\cal D}_{\alpha}
X^I)b\partial_z z^{\alpha}-X^I(\partial_z b-iA_z b)- \right.\nonumber \\
&&\left.({\cal D}_{\bar\alpha}\bar X^I)\bar b\partial_z\bar z^{\bar\alpha}-2ig_J
(\text{Im}\,{\cal N})^{-1|IJ}\right]\,. \label{fluxes}
\end{eqnarray}


\subsection{1/2 BPS near-horizon geometries}
\label{1/2BPS}

An interesting class of half-supersymmetric backgrounds was obtained in \cite{Klemm:2010mc}.
It includes the near-horizon geometry of extremal rotating black holes. The metric and the
fluxes read respectively
\begin{equation}
ds^2 = 4e^{-\xi}\left(-r^2dt^2 + \frac{dr^2}{r^2}\right) + 4(e^{-\xi} - Ke^{\xi})
(d\phi + r dt)^2 + \frac{4e^{-2\xi}d\xi^2}{Y^2(e^{-\xi} - Ke^{\xi})}\,, \label{near-hor}
\end{equation}
\begin{eqnarray}
F^I&=&16i\sqrt K\left(\frac{\bar X X^I}{1-iY}-\frac{X \bar X^I}{1+iY}\right)dt\wedge dr
\label{fluxes-1/2BPS} \\
&&+\frac{8\sqrt K}Y\left[\frac{2\bar X X^I}{1-iY}+\frac{2X\bar X^I}{1+iY}+\left(\mbox{Im}\,\mathcal{N}
\right)^{-1|IJ}g_J\right](d\phi + rdt)\wedge d\xi\ , \nonumber
\end{eqnarray}
where $X\equiv g_I X^I$, $K>0$ is a real integration constant, and $Y$ is defined by
\eq
Y^2=64e^{-\xi}|X|^2-1\ . \label{Y}
\feq
The moduli fields $z^{\alpha}$ depend on the horizon coordinate $\xi$ only, and obey the flow
equation\footnote{Note that this is not a radial flow, but a flow along the horizon.}
\eq
\frac{dz^{\alpha}}{d\xi} = \frac i{2\bar X Y}(1-iY)g^{\alpha\bar\beta}{\cal D}_{\bar\beta}\bar X\ .
\label{dzdxi}
\feq
\eqref{near-hor} is of the form (3.3) of \cite{Astefanesei:2006dd}, and describes the near-horizon
geometry of extremal rotating black holes\footnote{Metrics of the type \eqref{near-hor} were discussed
for the first time in \cite{Bardeen:1999px} in the context of the extremal Kerr throat geometry.},
with isometry group $\text{SL}(2,\bR)\times\text{U}(1)$.
From \eqref{dzdxi} it is clear that the scalar fields have
a nontrivial dependence on the horizon coordinate $\xi$ unless $g_I{\cal D}_{\alpha}X^I=0$.
As was shown in \cite{Klemm:2010mc}, the solution with constant scalars is the near-horizon
limit of the supersymmetric rotating hyperbolic black holes in minimal gauged
supergravity \cite{Caldarelli:1998hg}.

Using $Y$ in place of $\xi$ as a new variable, \eqref{dzdxi} becomes
\begin{equation}
\frac{dz^\alpha}{dY} = \frac{X g^{\alpha\bar\beta}{\cal D}_{\bar\beta}\bar X}{(Y-i)\left[-\bar X X +
{\cal D}_\gamma X g^{\gamma\bar\delta}{\cal D}_{\bar\delta}\bar X\right]}\,. \label{flow-Y}
\end{equation}
This can also be rewritten in a K\"ahler-covariant form, as a differential equation for the symplectic
section $\cal V$,
\begin{equation}
D_Y {\cal V} = \frac{X {\cal D}_\alpha {\cal V }g^{\alpha\bar\beta}{\cal D}_{\bar\beta}\bar X}{(Y-i)
\left[-\bar X X + {\cal D}_\gamma X g^{\gamma\bar\delta}{\cal D}_{\bar\delta}\bar X\right]}\,,
\end{equation}
where
\begin{equation}
D_Y \equiv \frac d{dY} + i A_Y
\end{equation}
denotes the K\"ahler-covariant derivative.

\subsection{The $\overline{\mathbb{C}\text{P}}^n$ model}
\label{CPn-model}

We shall now give an explicit example of a near-horizon
geometry with varying scalars, taking the $\overline{\mathbb{C}\text{P}}^n=\text{SU}(1,n)/(\text{SU}(n)
\times\text{U}(1))$ model, defined by the quadratic prepotential
\begin{equation}
F=\frac i4 X^I\eta_{IJ} X^J\,, \qquad \eta_{IJ} \equiv \text{diag}(-1,1,\ldots,1)\,.
\end{equation}
This yields
\begin{equation}
F_I = \frac{\partial F}{\partial X^I} = \frac i2\eta_{IJ} X^J\,.
\end{equation}
If we choose homogeneous coordinates by $Z^0=1$, $Z^\alpha=z^\alpha$, the holomorphic
symplectic section and the K\"ahler potential read respectively
\begin{equation}
v = \left(1, z^\alpha, -\frac i2, \frac i2 z^\alpha\right)^t\,, \qquad e^{-\cal K} = 1 -
\sum_{\alpha=1}^n |z^\alpha|^2\,,
\end{equation}
which implies that the complex scalars are constrained to the region $0\le\sum_\alpha|z^\alpha|^2<1$.
The special K\"ahler metric and its inverse are given by
\begin{equation}
g_{\alpha\bar\beta} = e^{\cal K}\delta_{\alpha\beta} + e^{2\cal K}{\bar z}^{\bar\alpha}z^\beta\,, \qquad
g^{\alpha\bar\beta} = e^{-\cal K}(\delta^{\alpha\beta} - z^\alpha {\bar z}^{\bar\beta})\,,
\end{equation}
while the period matrix is
\begin{equation}
{\cal N}_{IJ} = -\frac i2\eta_{IJ} + i\frac{Z_I Z_J}{Z_K Z^K}\,, \qquad \text{Im}\,{\cal N}_{IJ} = -\frac12
\eta_{IJ} + \frac12\left(\frac{Z_I Z_J}{Z_K Z^K} + \text{c.c.}\right)\,,
\end{equation}
\begin{equation}
(\text{Im}\,{\cal N})^{-1|IJ} = 2\left[-\eta^{IJ} + \left(\frac{Z^I\bar Z^J}{Z^K\bar Z_K} + \text{c.c.}\right)
\right]\,, \label{ImN-inverse}
\end{equation}
where we defined $Z_I\equiv\eta_{IJ}Z^J$. The scalar potential \eqref{scal-pot} reads
\begin{equation}
\label{eq:CPpotential}
V = 4 g^2 - 8\frac{\left|g_0 + \sum_\alpha g_\alpha z^\alpha\right|^2}{1 -
\sum_\beta |z^\beta|^2}\,,
\end{equation}
with $g^2\equiv\eta^{IJ}g_Ig_J$ from now on. $V$ has an extremum at $z^\alpha=-g_\alpha/g_0$,
where $V=12g^2$. For $z^\alpha=-g_\alpha/g_0$ to lie in the allowed region, the vector of gauge
couplings $g_I$ must be timelike, i.e., $g^2<0$. The extremum corresponds then to a
supersymmetric AdS vacuum. In addition, it is easy to see that the potential has flat directions given by
$g_0+\sum_\alpha g_\alpha z^\alpha=0$, where $V=4g^2$. For $n=1$, the flat directions
degenerate to the point $z^1=-g_0/g_1$, which lies in the allowed region for $g^2>0$. In this case
one has thus a critical point corresponding to a supersymmetry-breaking de~Sitter vacuum. If there
is more than one vector multiplet, the situation is of course more complicated.

\subsection{The $\text{t}^3$ model}
\label{t3-model}

Cubic models are of special interest. In the ungauged theory, these can be embedded in higher dimensional supergravity theories describing the low energy limit of some string theory. This appealing property is also displayed after gauging the theory at least for some of the cubic models. This is the case for the FI-gauged $stu$ model, which contains $n=3$ vector multiplets, and represents the best known example. It can be obtained as a consistent truncation of eleven-dimensional supergravity compactified on $\text{S}^7$ \cite{Cvetic:1999xp}. Moreover, if the three vector multiplets are identified, one gets the so-called
$\text{t}^3$ model, which we will consider in this work. The bosonic content of the theory contains the 
metric $g_{\mu\nu}$, two gauge fields $A^I_\mu$ and one complex scalar $\tau$. The theory is defined by 
the prepotential

\eq
F= -\frac{{(X^1)}^3}{X^0} \, .
\feq

\noindent 
In this case we have

\eq
F_I=X^0 \left(  \tau^3 , -3 \tau^2 \right) \, ,
\feq

\noindent
where we use homogeneous coordinates in the scalar manifold, with $\tau \equiv X^1/X^0$. The K\"ahler potential and scalar metric are then

\eq
e^ {-\mathcal{K}}=8\left( \text{Im} \tau \right)^3 \, , \qquad \qquad
g_{\tau \bar{\tau}}= \frac{3}{4 (\text{Im} \tau)^2} \, ,
\feq

\noindent
which implies $\text{Im} \tau > 0$. The scalar potential is then

\eq
V=-\frac{4g_1^2}{3 (\text{Im} \tau)} \, ,
\feq

\noindent
which has no critical point, so the theory does not admit AdS$_4$ vacua with constant moduli. Still, we will be able to construct a nontrivial family of black hole solutions, which of course do not asymptote to AdS$_4$.


\section{Supersymmetric rotating black holes in the $\overline{\mathbb{C}\text{P}}^n$ model}
\label{1/4BPS}

In this section we obtain a generalization of the asymptotically AdS black holes found in \cite{Klemm:2011xw, Colleoni:2012jq} to include an arbitrary number of vector multiplets $n$. To do so,
we shall use some ans\"atze which are inspired by those articles. We begin constructing in detail a
rotating black hole specified in terms of $n+2$ parameters -- see \eqref{CPPD}, \eqref{CPscalars} and 
\eqref{CPAI}. Moreover, in section \ref{sec-NUT} we present a solution with NUT charge.

\subsection{Solving the BPS equations}

A natural generalization of the successful ansatz used in \cite{Klemm:2011xw} for the $\overline{\mathbb{C}\text{P}}^1$ model is given by
\begin{equation}
\frac{\bar{X}^I}b = \frac{f^I(z) + \eta^I(w,\bar{w})}{\gamma(z)}\,, \qquad e^{2\Phi} = h(z)\ell(w, \bar{w})\,,
\end{equation}
where  $f^I(z)$ is a purely imaginary function, while $\gamma(z)$, $\eta^I(w,\bar{w})$, $h(z)$ and
$\ell(w,\bar{w})$ are real. With these assumptions, the BPS equations, although remaining nonlinear,
become separable and can be solved. The first of them, \eqref{dzPhi}, boils down to

\begin{equation}\label{primosempl}
\partial_z \mathrm{ln}h= - \frac{8 g_I \mathrm{Im}f^I}{\gamma}.
\end{equation}

\noindent
One can see that the symplectic constraint (\ref{sympconst}) implies that $X^I \eta_{IJ} \bar{X}^J=-1$,
which in turn gives
\begin{equation}
|b|^{-2} = \frac{1}{\gamma^2} \eta_{IJ} \left( f^I f^J - \eta^I \eta^J \right).
\end{equation}

\noindent
Using these expressions, equation (\ref{Delta-Phi}) reduces to
\begin{equation}
\frac{\partial \bar{\partial} \ln \ell}{\ell}= h \bigg[ - \frac{1}{4} \partial_z^2 \ln h + \frac{4}{\gamma^2} g_I g_J \bigg( \eta^{IJ} \eta_{LK} (\eta^L \eta^K- f^L f^K) + 2 (f^I f^J + \eta^I \eta^J) \bigg)    \bigg].
\end{equation}
Now we observe that if we take $h/\gamma^2=\text{const.}\equiv c_1>0$, this differential equation is separable, and one can define a constant $c_2$ such that 

\begin{align}\label{c1c2}
- \frac{h}{4} \partial_z^2 \ln h + \frac{4 h}{\gamma^2} g_I g_J \left( - \eta^{IJ} \eta_{LK} f^L f^K + 2 f^I f^J \right) &= c_1 c_2\,,
\\
\label{logl}
\frac{\partial \bar{\partial} \ln \ell}{\ell} - 4 c_1 g^2 \eta_K \eta^K - 8 c_1 \eta^2 &= c_1 c_2\,,
\end{align}

\noindent
where $\eta \equiv g_I \eta^I$ and capital indices are lowered with $\eta_{IJ}$. Equations \eqref{primosempl} and (\ref{c1c2}) can be solved by using the polynomial ansatz
\begin{equation}
\gamma= c + a z^2\,, \qquad h=c_1 (c+ a z^2)^2\,, \qquad f^I = i(\alpha^I z + \beta^I)\,,
\end{equation}

\noindent
for some real constants $a,c, \alpha^I, \beta^I$, which are constrained by
\begin{equation} \label{eq1}
g_I\alpha^I = -\frac a2\,, \quad g_I\beta^I = 0\,, \quad \alpha^I\eta_{IJ}\beta^J =0\,, \quad
-ac + 4 g^2\beta^2 = c_2\,, \quad a^2 = 4 g^2\alpha^2\,,
\end{equation}
where $\alpha^2\equiv\eta_{IJ}\alpha^I\alpha^J$ and $\beta^2\equiv\eta_{IJ}\beta^I\beta^J$.\\
The Bianchi identities (\ref{bianchi}) are then easily solved, and lead to
\begin{equation} \label{eq2}
\alpha^I = - \frac{2\eta^{IJ} g_J\alpha^2}a\,.
\end{equation}

\noindent
Observe that the set of constraints obtained so far completely fixes $\alpha^I$ and $c_2$ in terms of $a$, 
$c$ and $\beta^\alpha$, while $c_1$ remains free. As we will see, some of these degrees of freedom
can be eliminated by a coordinate transformation.\\
After some computation, Maxwell's equations (\ref{maxwell}) reduce to
\begin{equation} \label{maxw}
\partial\bar{\partial}\eta_I - 4\ell c_1\eta g_I\left(\eta_K\eta^K + \beta^2 - \frac{\alpha^2 c}a\right) = 0\,.
\end{equation}

\noindent
Together with \eqref{logl}, they define a system of $n+1$ second order, nonlinear differential equations, 
and looking for the general solution might seem a hopeless endeavour. Remarkably, the system can be solved using the ansatz of the type considered in \cite{Klemm:2011xw},
\begin{equation}
\label{CPansatzCS}
\ell = \frac{1+\delta}{\mathrm{cosh}^4(k\tilde{x})}\,, \quad
\eta^I = \hat{\eta}^I\mathrm{tanh}(k\tilde{x})\,, \quad
\delta = A\mathrm{cosh}^4(k\tilde{x})\,, \quad
\frac{dx}{d\tilde{x}} = \frac{\mathrm{cosh}^2 (k\tilde{x})}{1+\delta}\,,
\end{equation}

\noindent
where $A, k, \hat{\eta}^I $ are some constants and $ x \equiv(w+ \bar{w})/2$. Defining
$\hat\eta\equiv g_I\hat\eta^I$, equation (\ref{logl}) becomes
\begin{equation}
 k^2 + c_1 c_2 + \mathrm{sinh}^2 (k \tilde{x}) \left( - 2 k^2   + c_1 c_2 + 4 c_1 g^2 \hat{\eta}_K \hat{\eta}^K  + 8 c_1 \hat{\eta}^2 \right)=0 \,,
\end{equation}
which is solved provided
\begin{equation} \label{ris2}
k^2 = - c_1 c_2 \,, \qquad
3 k^2 = 4 c_1 g^2 \hat{\eta}_K \hat{\eta}^K + 8 c_1 \hat{\eta}^2 \,.
\end{equation}

\noindent
On the other hand, Maxwell's equations~\eqref{maxw} simplify to
\begin{equation}
k^2 \hat{\eta}_I + \mathrm{sinh}^2(k \tilde{x}) 4 c_1 \hat{\eta} g_I \left( \hat{\eta}_K \hat{\eta}^K + \beta^2 - \frac{\alpha^2 c}a\right) + 4c_1 \hat{\eta} g_I \left( \beta^2 - \frac{\alpha^2 c}a\right) = 0\,,
\end{equation}

\noindent
which are satisfied if
\begin{equation} \label{ris1}
\hat{\eta}^K \hat{\eta}_K + \beta^2 - \frac{\alpha^2 c}a = 0 \,, \qquad
k^2 \hat{\eta}_I + 4 c_1 \hat{\eta} g_I \left( \beta^2 - \frac{\alpha^2 c}a\right) = 0\,.
\end{equation}

\noindent
In summary, we can combine~\eqref{eq1}, \eqref{ris2} and~\eqref{ris1} to find
\begin{equation}
k^2= 4 c_1 \hat{\eta}^2\,, \qquad g^2 \hat{\eta}_I = \hat{\eta} g_I\,.
\end{equation}

\noindent
This implies that the only independent parameter in~\eqref{CPansatzCS} is $A$.\\
Finally, to completely specify the solution we have to integrate (\ref{dsigma}). To this end we use
$(\tilde{x}, y, z)$ as coordinates, where $y=(w- \bar{w})/2i$. The relevant Hodge duals on the metric 
\eqref{metr-base} are
\begin{equation*}
\star^{(3)} d\tilde{x} = \frac{1+\delta}{\mathrm{cosh}^2(k \tilde{x})} dy\wedge dz\,, \qquad
\star^{(3)} dz = \frac{e^{2\Phi}\mathrm{cosh}^2 (k\tilde{x})}{1+\delta} d\tilde{x}\wedge dy\,,
\end{equation*}
and thus (\ref{dsigma}) takes the form 
\begin{equation}
\begin{split}
&\partial_{\tilde{x}} \sigma_y d \tilde{x} \wedge dy - \partial_z \sigma_y dy \wedge dz - \frac{k}{\gamma^2} \left( \alpha^I \hat{\eta}_I z \right) \frac{1+\delta}{\mathrm{cosh}^4(k \tilde{x})} dy \wedge dz - \frac{2 \hat{\eta} \alpha^2 c_1}{a} \frac{\mathrm{tanh}(k \tilde{x})}{\mathrm{cosh}^2(k \tilde{x})} d \tilde{x} \wedge dy \\
& + 4 c_1 \hat{\eta} \left( \alpha^2 z^2 + \beta^2 + \hat{\eta}_K \hat{\eta}^K \mathrm{tanh}^2(k \tilde{x})  \right) \frac{\mathrm{tanh}(k \tilde{x})}{\gamma \mathrm{cosh}^2(k \tilde{x})} d \tilde{x} \wedge dy =0\,,
\end{split}
\end{equation}
which can be easily integrated to give
\begin{equation} \label{sigma}
\sigma_y = \frac{\hat{\eta}}{4 g^2 k}\left[\frac{a c_1}{\mathrm{cosh}^2(k \tilde{x})} - \frac{k^2}{\gamma} \left( A + \frac{1}{\mathrm{cosh}^4(k \tilde{x})}\right)\right]\,.
\end{equation}

\subsection{The fields}

The metric \eqref{gen-metr} of the solution obtained here can be simplified by the coordinate transformation
\begin{equation}\label{ruoto}
\left( \begin{array}{c}
t\\
y
\end{array}\right)\mapsto
\sqrt{\frac{\mathrm{\textsf{E}}}{-2 A}}
\left( \begin{array}{cc}
0 & - \frac{a A \mathrm{\textsf{E}} L^3}{8 \hat{\eta}} \\
-\frac{1}{k L} & \frac{\mathrm{\textsf{E}} L}{2k}
\end{array} \right)
\left( \begin{array}{c}
t\\
y
\end{array} \right) \, ,
\qquad
p = B \mathrm{tanh}(k \tilde{x}) \, ,
\qquad
q = D z \, ,
\end{equation}

\noindent
where $B = \sqrt{\frac{\mathrm{\textsf{E}}}{-8 g^2}}$, $D = \sqrt{\frac{a^2 c_1 \mathrm{\textsf{E}}}{- 8 g^2 k^2}}$ and $\mathrm{\textsf{E}}$ is a positive constant, so $A$ must be negative. In these coordinates the metric takes the Carter-Pleba\'nski \cite{Carter:1968ks,Plebanski:1975xfb} form\footnote{Notice that
the metrics \eqref{CPPD} and \eqref{nonextr-metric}, though looking very similar to the
Carter-Pleba\'nski form, are not contained in this family since they are not of Petrov type D.
Rather, they belong to the more general class called the Benenti-Francaviglia metric \cite{Benenti:1979},
which admits a Killing tensor. We would like to thank M.~Nozawa for pointing out this to us.}
\begin{equation} \label{CPPD}
\begin{split}
& ds^2 =\frac{p^2 + q^2 - \Delta^2}{P} dp^2 + \frac{P}{p^2+q^2 -\Delta^2} \left( dt +(q^2- \Delta^2)dy \right)^2+\\
& + \frac{p^2 +q^2 -\Delta^2}{Q} dq^2 - \frac{Q}{p^2+q^2-\Delta^2} \left( dt- p^2 dy \right)^2 \, ,
\end{split}
\end{equation}

\noindent
where
\begin{equation}\label{PQ}
P = (1+A) \frac{\mathrm{\textsf{E}}^2 L^2}{4} - \mathrm{\textsf{E}} p^2 + \frac{p^4}{L^2}\,, \qquad
Q = \frac{1}{L^2} \left(q^2 + \frac{\mathrm{\textsf{E}} L^2}{2}- \Delta^2 \right)^2\,,
\end{equation}
and $L^2$ and $\Delta^2$ are two positive constants defined by
\begin{equation}
\Delta^2 =\frac{\mathrm{\textsf{E}} \beta^2}{8 \hat{\eta}^2}\,, \qquad L^2 = - \frac1{4 g^2}\,.
\end{equation}
The scalar fields $z^\alpha$ read
\begin{equation*}
z^\alpha = -\frac{g_\alpha}{g_0}\frac{p^2 + q^2 + i\Delta_1 p - \Delta_1 q}{p^2 + q_1^2} - i\Delta_1
\frac{\beta^\alpha}{\beta^0}\frac{p - i q_1}{p^2 + q_1 ^2}\,,
\end{equation*}
or equivalently, in a neater fashion
\begin{equation}\label{CPscalars}
z^\alpha =\frac{1}{p + i q_1} \left(-\frac{g_\alpha}{g_0} (p + i q) - i\Delta_1
\frac{\beta^\alpha}{\beta^0} \right),
\end{equation}
with
\begin{equation}
q_1 = q- \Delta_1\,, \qquad\Delta_1 = \frac{\beta^0}{g_0}\sqrt{\frac{-g^2}{\beta^2}}\Delta\,.
\end{equation}

\noindent
If $\Delta_1=0$ (or equivalently $\Delta=0$), the scalars are constant and they assume the value
$-g_\alpha/g_0$ for which the potential (\ref{eq:CPpotential}) is extremized.\\
To complete the solution we need an expression for the gauge potentials, which are found by
integrating (\ref{fluxes}). This leads to
\begin{equation} \label{CPAI}
A^I = 2 \eta^{IJ} g_J \mathrm{\textsf{E}} L^2 \sqrt{-A} \frac{p}{p^2 +q^2 -\Delta^2} \left( dt + (q^2 -\Delta^2) dy \right).
\end{equation}
The solution is thus specified by $n+2$ free real parameters, and therefore represents a generalization
of the black holes with $n=1$ constructed in \cite{Klemm:2011xw}. The parameters can be taken to be
$A$, $\textsf{E}$, $\Delta$ and $\beta^\alpha/\beta^0$, subject to the constraint $g_I\beta^I=0$,
cf.~\eqref{eq1}.\\
A particular, interesting choice is given by
\begin{equation}
\sqrt{-A} = \frac{L^2 + j^2}{L^2 - j^2}\,, \qquad \mathrm{\textsf{E}} = \frac{j^2}{L^2} -1\,.
\end{equation}

\noindent
Then, after the change of coordinates
\begin{equation}
p = j\cosh\theta\,, \qquad y = -\frac{\phi}{j\Xi}\,, \qquad t = \frac{T - j\phi}{\Xi}\,, \qquad
\Xi\equiv 1 + \frac{j^2}{L^2}\,,
\end{equation}
and defining the functions
\begin{equation}
\rho^2= q^2 + j^2 \cosh^2 \theta, \ \ \ \Delta_q =\frac{1}{L^2} \left( q^2 + \frac{j^2-L^2}{2} - \Delta^2\right)^2, \ \ \ \Delta_\theta = 1 + \frac{j^2}{L^2}\cosh^2 \theta,
\end{equation}

\noindent
the metric~\eqref{CPPD}, the scalars~\eqref{CPscalars} and the gauge potentials~\eqref{CPAI} become
respectively
\begin{align}
\nonumber
& ds^2 = \frac{\rho^2 - \Delta^2}{\Delta_q} dq^2 + \frac{\rho^2 - \Delta^2}{\Delta_\theta} d \theta^2 + \frac{\Delta_\theta \sinh^2\!\theta}{(\rho^2 -\Delta^2)\Xi^2} \left( j dT - (q^2+ j^2- \Delta^2) d \phi \right)^2  \\
\label{metricj}
& \quad -\frac{\Delta_q}{(\rho^2 - \Delta^2) \Xi^2} \left( dT + j \sinh^2\!\theta d \phi \right)^2 ,
\\
\nonumber
& z^\alpha = \frac1{j^2 \cosh^2 \theta + q_1^2}\left[ - i \Delta_1\left(\frac{g_\alpha}{g_0}
\left(j\cosh \theta + i q \right) + \frac{\beta^\alpha}{\beta^0}\left(j\cosh\theta -i q\right)\right)\right. \\
\label{scalarj}
&\quad\left. - \frac{g_\alpha}{g_0} \rho^2 + \frac{\beta^\alpha}{\beta^0} \Delta_1^2 \right], \\
& A^I = 2\eta^{IJ} g_J L^2 \frac{\cosh\theta}{\rho^2 - \Delta^2} \left( j dT - \left( q^2 + j^2 - \Delta^2 \right) d\phi\right).
\end{align}

\noindent
The metric depends only on the two constants $\Delta$ and $j$, that can be interpreted respectively
as scalar hair and rotation parameters. Note that for $j=0$ the scalars are real, whereas in the rotating
case there is a nontrivial axion.

\subsection{Near-horizon limit}

The metric (\ref{metricj}) has an event horizon at $\Delta_q=0$, i.e., for $q=q_h$ with
\begin{equation}
q_{\mathrm{h}}^2 =\Delta^2 + \frac{1}{2} \left( L^2 - j^2\right).
\end{equation}
To obtain the near-horizon geometry, we set
\begin{equation}
q = q_{\mathrm{h}} + \epsilon q_0 z\,, \qquad T = \frac{\hat{t} q_0}{\epsilon}\,, \qquad \phi =
\hat{\phi} + \Omega\frac{\hat{t} q_0}{\epsilon}\,,
\end{equation}
and then zoom in by taking the limit $\epsilon\rightarrow 0$. The parameter
$\Omega = j/(q_{\mathrm{h}}^2 + j^2 - \Delta^2)$ represents the angular velocity of the horizon,
while $q_0\equiv\frac{L^2 \Xi}{2 \sqrt{2} q_{\mathrm{h}}}$. In this limit the metric boils down to
\begin{equation}
\begin{split}
&ds^2 = \frac{\rho^2_\mathrm{h} - \Delta^2}{4 q_{\mathrm{h}}^2 z^2 } L^2 dz^2 + \frac{\rho^2_\mathrm{h} - \Delta^2}{\Delta_\theta} d\theta^2 + \frac{L^4 \Delta_\theta \sinh^2\!\theta}{4 (\rho^2 _\mathrm{h} - \Delta^2)} \left( d \hat{\phi}+ \frac{j}{q_{\mathrm{h}}} z d \hat{t}\right)^2\\
&\qquad -\frac{\rho^2_{\mathrm{h}}- \Delta^2 }{4 q_{\mathrm{h}}^2} L^2 z^2 d\hat{t}^2,
\end{split}
\end{equation}
where $\rho^2_\mathrm{h} \equiv q_{\mathrm{h}}^2 + j^2 \cosh^2\!\theta$.\\ 
The final coordinate transformation
\begin{equation}
e^{-\xi} = L^2 \frac{q_{\mathrm{h}}^2 + j^2 \cosh^2\!\theta - \Delta^2}{16 q_{\mathrm{h}}^2}\,,
\qquad x = \frac{q_{\mathrm{h}}}j\hat{\phi}\,,
\end{equation}
casts the metric into the form \eqref{near-hor}, namely
\begin{equation}\label{NHmetric}
ds^2 = 4 e^{- \xi} \left( - z^2 d \hat{t}^2 + \frac{dz^2}{z^2}\right) + 4\left(e^{-\xi} - K e^\xi\right)
\left(dx + z d\hat{t}\right)^2 + \frac{4 e^{-2 \xi} d \xi^2}{Y^2 (e^{-\xi} - K e^\xi)}\,,
\end{equation}
where $K \equiv L^8 \Xi^2 /1024 q_{\mathrm{h}}^4$.\\
There is thus a supersymmetry enhancement for the near-horizon geometry, which preserves half of the 8 
supercharges of the theory. Exploiting this fact, there is an alternative way to arrive at this solution that
goes as follows. In the $\overline{\mathbb{C}\text{P}}^n$ model, the flow equation \eqref{flow-Y} becomes
\begin{equation}
\frac{dz^\alpha}{dY} = \frac{(g_\alpha + z^\alpha g_0)(g_0 + \sum_\beta g_\beta z^\beta)}{g^2(Y-i)}\,,
\end{equation}
which is solved by
\begin{equation}
z^\alpha = \frac{\mu_\alpha g_0 - g_\alpha (Y-i)}{g_0(Y - i - \mu_0)}\,. \label{scalars-near-hor-CPn}
\end{equation}
Here, $\mu_I=(\mu_0,\mu_\alpha)\in\mathbb{C}^{n+1}$ is a constant vector orthogonal to the gauge coupling, $\mu_I\eta^{IJ}g_J=0$. One can now compute $|X|^2$ as a function of $Y$, with the result
\begin{equation}
|X|^2 = -\frac{g^4(Y^2 + 1)}{g^2(Y^2 + 1) + g_0^2\mu\cdot\bar\mu}\,, \qquad \text{where} \, \, \,  \, 
\mu\cdot\bar\mu = \mu_I\eta^{IJ}\bar\mu_J\,.
\end{equation}
Plugging this into \eqref{Y} gives then $\xi(Y)$, and the metric \eqref{near-hor} becomes
\begin{eqnarray}
ds^2 &=& \frac{-g^2(Y^2+1) - g_0^2\mu\cdot\bar\mu}{16 g^4}\left(-r^2 dt^2 + \frac{dr^2}{r^2}\right)
+ \frac{P(Y)(d\phi + rdt)^2}{16g^4[-g^2(Y^2+1) - g_0^2\mu\cdot\bar\mu]} \nonumber \\
&& + \frac{[-g^2(Y^2+1) - g_0^2\mu\cdot\bar\mu] dY^2}{4 P(Y)}\,, \label{metr-near-hor-CPn}
\end{eqnarray}
where we defined the quartic polynomial
\begin{equation}
P(Y) = [g^2(Y^2 + 1) + g_0^2\mu\cdot\bar\mu]^2 - K (64g^4)^2\,.
\end{equation}
Using $Y=-j\cosh\theta/q_{\mathrm{h}}$, one finds that the modulus of the parameters $\mu_I$ is
related to $\Delta$ by
\begin{equation*}
\mu \cdot \bar{\mu} = \frac{\Delta^2}{4 L^2 g_0^2 q_{\mathrm{h}}^2}\,.
\end{equation*}
The expression for the vector $\mu_I$ can be found requiring that the scalar fields (\ref{scalarj}) coincide in the near-horizon limit with the expression (\ref{scalars-near-hor-CPn}), yielding
\begin{equation}
\mu_0 = -i\frac{\Delta_1}{q_{\mathrm{h}}}\,, \qquad \mu_\alpha =
i\frac{\Delta_1\beta^\alpha}{q_{\mathrm{h}}\beta^0}\,.
\end{equation}
Since the static supersymmetric black holes in the $\overline{\mathbb{C}\text{P}}^n$ model constructed
in \cite{Cacciatori:2009iz} have necessarily hyperbolic horizons, one may ask whether spherical rotating horizons
are possible. As was discussed in detail in \cite{Klemm:2014nka}, this
question is intimately related to the behaviour of $P(Y)$. Namely, for spherical horizons to be feasible $P(Y)$ must have
four distinct roots, and then $Y$ is restricted to the region between the two central roots where $P(Y)$
is positive. The latter condition, together with $-g^2(Y^2+1) - g_0^2\mu\cdot\bar\mu>0$, is necessary in order for the metric to have the correct signature\footnote{Note that there is a curvature
singularity for $-g^2(Y^2+1) - g_0^2\mu\cdot\bar\mu=0$.}. Imposing $P(Y)=0$ yields
\begin{equation}
-g^2(Y^2+1) - g_0^2\mu\cdot\bar\mu = 64g^4\sqrt K\,.
\end{equation}
There are thus only two roots $\pm Y_0$
(with $Y_0>0$), and spherical horizons are therefore excluded in the rotating solution as well. One can show that, in the static limit, the near-horizon geometry of the black holes constructed in \cite{Cacciatori:2009iz} is recovered.

\subsection{NUT-charged black holes}
\label{sec-NUT}
In this section we construct supersymmetric NUT-charged black holes. To do so it is sufficient to mimic
what was done in \cite{Colleoni:2012jq}, where the theory with only one vector multiplet was considered.
Since the BPS equations can be solved following the same steps of that paper, we will just briefly summarize the process here and refer to \cite{Colleoni:2012jq} for further details. We assume that both the
scalars and the function $b$ depend on the coordinate $z$ only, and use the ansatz
\begin{equation}
\frac{X^I}{\bar{b}}= \frac{\alpha^I z + \beta^I}{z^2 + i D z + C}\,, \qquad
\Phi = \psi (z) + \gamma (w, \bar{w})\,,
\end{equation}
where $\alpha^I,\beta^I,C$ are complex constants and $D$ is a real constant. The dependence of the solution on the coordinates $w,\bar w$ is obtained from (\ref{Delta-Phi}), which reduces to
\begin{equation} \label{kappa}
- 4 \partial \bar{\partial} \gamma = \kappa e^{2 \gamma} \, ,
\end{equation}
where $\kappa$ is a constant whose value will be fixed later. This is the Liouville equation for the metric
$e^{2\gamma}dw d\bar{w}$, which consequently has constant curvature $\kappa$. We will take as a particular solution
\begin{equation}
e^{2\gamma} = \left(1 + \frac{\kappa}4 w\bar{w}\right)^{-2}.
\end{equation}
From (\ref{dzPhi}) one gets
\begin{equation}
\psi(z) = \text{Im}\alpha\left(\ln[ z^4 + z^2 (2\text{Re} C + D^2) + 2 D z\text{Im}C + |C|^2]\right)\,,
\end{equation}
provided the following constraints are satisfied
\begin{equation}\label{rel1}
\text{Im}\beta = D\text{Re}\alpha\,, \quad -2[\text{Im}(\bar{\alpha} C) + D\text{Re}\beta] = \text{Im}
\alpha (2\text{Re} C + D^2)\,, \quad -2\text{Im}(\bar{\beta} C) = D\text{Im}\alpha\text{Im} C\,,
\end{equation}
where $\alpha\equiv g_I\alpha^I$ and $\beta\equiv g_I\beta^I$.\\
The expressions displayed so far coincide with those found for the case with one vector multiplet; the explicit form of the prepotential has not been used to solve (\ref{dzPhi}) and (\ref{Delta-Phi}). In order to solve the remaining BPS equations we choose $\mathrm{Im} \alpha= 1/2$, since in this way they assume a
polynomial form. Then the Bianchi identities fix $\alpha^I$ and $\kappa$ to
\begin{equation}
\label{curvatura}
\alpha^I = \frac i{2g^2}\eta^{IJ} g_J\,, \qquad \kappa = 2\text{Re} C - 8 g^2\beta^K\bar{\beta}_K\,.
\end{equation}
On the other hand, Maxwell's equations are automatically satisfied provided these relations hold. Finally, 
integration of (\ref{dsigma}) gives
\begin{equation}
\sigma = i \frac{D}{32 g^2} \frac{ \bar{w}dw-w d \bar{w}}{1 + \frac{\kappa}{4} w \bar{w}}\,,
\end{equation}
from which it is evident that the parameter $D$ is related to the NUT charge of the solution.\\
The warp factor of the metric is in this case
\begin{equation}
|b|^{-2} = - \frac{ z^2 +4g^2\beta^K \bar{\beta}_K  }{4 g^2 |z^2 + i D z + C|^2}\,,
\end{equation}
where we recall that in this model $g^2<0$. The solution will have an event horizon at $z=z_{\text h}$
if $b(z_{\text h})$ vanishes, which happens for
\begin{equation}
z_{\text h}^2 = -\text{Re} C\,, \qquad D z_{\text h} = -\text{Im}C\,.
\end{equation}
This is possible if $(\text{Im} C)^2 = - D^2\text{Re} C$ and $\text{Re} C<0$. There is a curvature
singularity at $z^2+ 4 g^2 \beta^K \bar{\beta}_K=0$, which is hidden behind the horizon if
\begin{equation}\label{condiz}
\text{Re} C < 4 g^2 \beta^K \bar{\beta}_K.
\end{equation}
Then, from (\ref{curvatura}) we see that $\kappa <0$ and therefore the horizon is always hyperbolic.\\
The solution is in principle specified by $2n+2$ real parameters, which can be taken as $\beta^I$, $D$ and
$\text{Re} C$ with the constraint $\beta=-D/4$, which follows from \eqref{rel1}. If \eqref{condiz}
holds, the metric describes a regular black hole. Notice that we can use the scaling symmetry
$(t,z,w,C,D,\beta^I,\kappa)\mapsto (t/\lambda,\lambda z, w/\lambda,\lambda^2 C,\lambda D,\lambda 
\beta^I,\lambda^2\kappa)$ to set $\kappa=-1$ without loss of generality, which reduces the number
of independent parameters to $2n+1$.\\
The fluxes can be computed by plugging the results found so far into (\ref{fluxes}). A long but 
straightforward calculation yields
\begin{equation}\begin{split} \label{flussonut}
F^I &= 4(dt + \sigma)\wedge dz\frac1{(z^2 + 4 g^2\beta^K\bar{\beta}_K)^2}\bigg[4 g^2\left(2\text{Im}
C\text{Im}\beta^I - \text{Re}\beta^I\right) z \\
& - 2\eta^{IJ} g_J D z (1 + 2\text{Re} C) + \left(-1 - 2\text{Re} C + 2 z^2\right)\left(2 g^2 D\text{Im}
\beta^I + \eta^{IJ} g_J\text{Im} C\right)\bigg] \\
& -\frac12 e^{2\gamma} dw\wedge d\bar{w}\frac i{4 g^2 (z^2 + 4 g^2\beta^K\bar{\beta}_K)}\bigg[
\eta^{IJ} g_J\left( -1 - 2(\text{Re} C + z^2)\right) \\
& + 4 D\eta^{IJ} g_J z\left(D z + \text{Im} C\right) + 8 g^2 D\left(\text{Re}\beta^I z^2 + D\text{Im}\beta^I
z + \text{Re}(\bar{C}\beta^I)\right)\bigg].
\end{split}
\end{equation}
The magnetic and electric charges of the solution are given by
\begin{equation}
P^I = \frac1{4\pi}\int_{\Sigma_\infty} F^I\,, \qquad Q_I = \frac1{4\pi}\int_{\Sigma_\infty} G_I\,,
\end{equation} 
where $\Sigma_\infty$ denotes a surface of constant $t$ and $z$ for $z\rightarrow\infty$, and $G_I$ is 
obtained from the action as $G_I=\delta S/\delta F^I$. This leads to
\begin{equation}\label{magnetica}
\frac{P^I}V = \frac{1 - 2 D^2}{8\pi g^2}\eta^{IJ} g_J  - \frac{D\text{Re}\beta^I}{2\pi}\,, \qquad
\frac{Q_I}V = -\frac{ g_I\text{Im} C}{8\pi g^2} + \frac{\eta_{IJ}\text{Im}\beta^J D}{4\pi}\,,
\end{equation}
where $V$ is defined by
\begin{equation}
V= \frac{i}{2} \int e^{2 \gamma} dw \wedge d \bar{w}\,.
\end{equation}
Finally, the scalars read
\begin{equation}
z^\alpha = \frac{2g^2\beta^\alpha + i g_\alpha z }{2g^2 \beta^0- i g_0 z }\,.
\end{equation}

\section{Supersymmetric rotating black holes in the $\text{t}^3$ model}

\subsection{A near-horizon solution}\label{subsec:nh-t^3}

Before starting, we notice that when looking for solutions of the $\overline{\mathbb{C}\text{P}}^n$
model, it proved useful to work with a factorized ansatz for the real and imaginary components of $\bar{X}^ I/b$. If a similar decomposition is performed in the case at hand, the equations of motion do not factorize unless we assume that the real or imaginary part of $\bar{X}^ 0/b$ vanishes. We will only explore here the latter possibility, as in the former we just found trivial solutions. Since in homogeneous coordinates $X^0$ is purely real, one can see that this is equivalent to setting $\bar{b}=b$. We will thus use the ansatz
\eq
\label{eq:t3ansatz1}
\frac{\bar{X}^0}{b}=\frac{\eta^0 (w,\bar{w})}{\gamma (z)}\,, \qquad 
\frac{\bar{X}^1}{b}=\frac{f^1(z)+\eta^1 (w,\bar{w})}{\gamma (z)}\,, \qquad
e^{2\Phi}=h(z)\ell(w,\bar{w})
\feq
in the system of BPS equations \eqref{dzPhi}-\eqref{dsigma}. From \eqref{dzPhi} and \eqref{Delta-Phi}
we get
\eq \label{eq-f}
\partial_z \ln h = 8i \frac{g_1 f^1}{\gamma} \,, \qquad \qquad
\frac{\partial \bar\partial \ln\ell}{\ell} = -\frac14 \partial^2_z h - \frac{32}{3} \frac{h}{\gamma^2} ({g_1 f^1})^2\,.
\feq
Using the first equation, we find that the second is separable and boils down to
\eq
\label{eq:t3lh}
\dfrac{\partial \bar\partial \ln\ell}{\ell} = \dfrac{C_1}4\,, \qquad
\partial^2_z h -\dfrac23 \dfrac{(\partial_z h)^2}{h} = -C_1\,,
\feq
for some constant $C_1$. \eqref{eq:t3lh} determines the dependence on $w,\bar w$ and $z$ of the
three-dimensional base space. For $C_1\neq 0$\footnote{The case $C_1=0$ belongs to a qualitatively different family of solutions to \eqref{eq:t3lh}, which however does not seem to be well-suited for solving
the remaining differential equations of the system.} the solution for $h$ reads\footnote{$a$ and $c$ are
integration constants. Although $h$ and $f^1/\gamma$ depend only on the ratio $c/a$, we prefer to
keep them both, for reasons that become clear further below.}
\eq\label{eq:t3solh}
h(z) = \frac32 C_1 \left(z + \frac{c}{a}\right)^2\,,
\feq
which implies
\eq
\frac{f^1}{\gamma} = -\frac{i}{4 g_1 \left(z + \frac ca\right)}\,.
\feq
The first of \eqref{eq:t3lh} is just Liouville's equation, and thus the explicit form of $l(w, \bar{w})$
depends on the choice of a meromorphic function. In order to make further progress, from now on we
shall consider a particular case that has been proven successful for our purpose, i.e.,
\eq
l(w,\bar{w})=\frac2{C_1\sinh^2\!\left(\frac{w+\bar w}2\right)}\,.
\feq
Then the Bianchi identities \eqref{bianchi} are automatically solved, so that Maxwell's equations \eqref{maxwell} represent the last obstacle. Setting, like in the $\overline{\mathbb{C}\text P}^n$ case, $h(z)/\gamma(z)^2$ to a constant, the latter assume a simple form. The value of this constant is totally arbitrary, but with a redefinition of $a$, and thus of $c$ in order to keep $c/a$ unchanged, we can always bring it to $\frac{3C_1}{2a^2}$, in which case the Maxwell equations become
\eq\label{eq:Maxw-nh-t^3}
\partial\bar\partial \left[\dfrac{1}{{\eta^0}^2} - 48\dfrac{g_1^2}{a^2} R^2\right] = 0\,, \qquad
2\partial\bar\partial R - \dfrac{R}{\sinh^2\!\left(\frac{w+\bar w}2\right)} = 0\,,
\feq
where
\eq
R(w,\bar w)\equiv\frac{g_I\eta^I}{g_1\eta^0}\,.
\feq
The second equation of \eqref{eq:Maxw-nh-t^3} can be readily solved,
\eq\label{eq:R(x)-t^3}
R(x) = \Xi_1 \coth x + \Xi_2 \left[x \coth x - 1\right]\,,
\feq
where $\Xi_{1,2}$ are integration constants and $x\equiv(w+\bar w)/2$. The first of
\eqref{eq:Maxw-nh-t^3} implies
\eq
\frac{1}{{\eta^0}^2} = 48 \frac{g_1^2}{a^2} R^2 + \mathrm{Re} F(w)\,,
\feq
for some arbitrary function $F(w)$ that in a first step we will simply set to 0.\\
The equation \eqref{dsigma} for the shift vector $\sigma$ boils down to
\begin{displaymath}
\partial_z\sigma_w = -\frac{3i\partial R}{4g_1^2\left(z + \frac ca\right)^2}\,, \qquad
\partial_z\sigma_{\bar w} = \frac{3i\bar\partial R}{4g_1^2\left(z + \frac ca\right)^2}\,, \qquad
\partial\sigma_{\bar w} - \bar\partial\sigma_w = -\frac{3i\partial\bar\partial R}{2g_1^2\left(z +
\frac ca\right)}\,,
\end{displaymath}
which is solved by
\eq
\sigma = \frac{3i}{4g_1^2\left(z + \frac ca\right)}(\partial R dw - \bar\partial R d\bar w)\,.
\feq
Defining $y\equiv (w-\bar w)/(2i)$, the metric \eqref{gen-metr} becomes
\eq\label{metric}
ds^2 = -\frac{8 g_1^2}{\sqrt3 R}\left[\left(z + \frac ca\right) dt - \frac3{4g_1^2}\partial_x R dy\right]^2
+ \frac{\sqrt3 R}{2 g_1^2}\left[\frac{dz^2}{\left(z + \frac ca\right)^2} +
\frac{3(dx^2 + dy^2)}{\sinh^2\! x}\right],
\feq
while the scalar field is given by
\eq\label{scalar}
\tau  = -\frac{g_0}{g_1} + R(x) + i\sqrt{3} R(x) = -\frac{g_0}{g_1} + 2 e^{i \pi/3} R(x) \, .
\feq
For $\Xi_2=0$ one can readily identify this solution as belonging to the class of half-supersymmetric
near-horizon backgrounds presented in section \ref{1/2BPS}. Performing the change of coordinates
\eq
e^{-\xi} = \sqrt K\coth x\,, \qquad r = z + \frac ca\,, \qquad \phi = -\sqrt3 y\,, \qquad T =
\frac t{2\sqrt K}\,,
\feq
where $\sqrt K=\dfrac{\sqrt{3}\Xi_1}{8 g_1^2}$, the metric is brought to the form \eqref{near-hor} with 
$Y^2=1/3$, namely
\begin{equation}
\label{eq:t3nh}
ds^2 = 4e^{-\xi}\left(-r^2 dT^2 + \frac{dr^2}{r^2}\right) + 4(e^{-\xi} - K e^{\xi})
(d\phi + r dT)^2 + \frac{12 e^{-2\xi} d\xi^2}{e^{-\xi} - K e^{\xi}}\,.
\end{equation}
In the same way, one can check that the scalar \eqref{scalar} satisfies the flow equation \eqref{dzdxi}.

\subsection{Black hole extension}

We will now construct a black hole whose near-horizon geometry is given by the solution found in
the previous subsection. This is achieved with a slight generalization of the ansatz \eqref{eq:t3ansatz1}.
We maintain the factorization form of $e^{2\Phi}$ and Im$(\bar{X}^I/b)$, but leave Re$(\bar{X}^I/b)$
as arbitrary functions of the three spatial coordinates.\\
The first steps of subsection \ref{subsec:nh-t^3} that determine the functions $h(z)$, $l(w,\bar{w})$
and Im$(\bar{X}^I/b)$ remain identical. The difference appears in the first of Maxwell's equations, which
now read
\eq\label{eq:Maxw-bh-t^3-1}
-\frac{r^2}{\sinh^2\!\left(\frac{w + \bar w}2\right)}\partial_r^2\left[\frac1{r^2\text{Re}^2
\!\left({\bar X}^0/b\right)}\right] - \frac43\partial\bar\partial\left[\frac1{r^2\text{Re}^2
\!\left({\bar X}^0/b\right)}\right] + 64 g_1^2\partial\bar\partial\left(R^2\right) = 0\,,
\feq
\eq\label{eq:Maxw-bh-t^3-2}
2\partial \bar\partial R - \dfrac{R}{\sinh^2\!\left( \frac{w+\bar{w}}{2} \right)}  = 0\,,
\feq
where $r=z+c/a$. A simple solution to \eqref{eq:Maxw-bh-t^3-1} is
\eq
\text{Re}\!\left({\bar X}^0/b\right) = \frac1{4\sqrt3 g_1 r\sqrt{\alpha r + \beta + R(w,\bar w)^2}}\,,
\feq 
while \eqref{eq:Maxw-bh-t^3-2} is solved by \eqref{eq:R(x)-t^3}. Here, $\alpha$ and $\beta$ denote
integration constants. Then, the scalar, metric and gauge potentials read respectively
\eq
\tau = -\frac{g_0}{g_1} + R + i\sqrt{3} \sqrt{\alpha r + \beta + R^2} \,,
\feq
\eq \label{metric-z-x}
\begin{aligned}
ds^2 = & -\frac{8 g_1^2}{\sqrt3\sqrt{\alpha r + \beta + R^2}} \left[r \, dt + \frac{3}{4g_1^2} \partial_x R \, dy\right]^2 + \\
       & + \frac{\sqrt{3}}{2 g_1^2} \sqrt{\alpha r + \beta + R^2} \left[\frac{dr^2}{r^2} +
\frac{3(dx^2 + dy^2)}{\sinh^2\!x}\right]\,,
\end{aligned}
\feq
\eq
\begin{aligned}
A^0 & = -\frac{2g_1}{3(\alpha r + \beta + R^2)}\left(r dt + \frac3{4g_1^2}\partial_x R dy\right)\,, \\
A^1 & = -\frac{2g_1}{3(\alpha r + \beta + R^2)}\left(R - \frac{g_0}{g_1}\right)\left(r dt + \frac3{4g_1^2}
\partial_x R dy\right) - \frac{\coth x}{2g_1} dy\,.
\end{aligned}
\feq
Now the scalar depends on the radial coordinate as well, and we recover the near-horizon 
geometry discussed above by rescaling $r\mapsto\epsilon r$, $t\mapsto t/\epsilon$ and taking the limit $\epsilon \rightarrow 0$.\\
As we already mentioned, the asymptotic limit of this solution cannot be AdS$_4$ since the scalar
potential has no critical points. For large values of $r$, the metric behaves as
\eq
ds^2 = d\rho^2 + \frac{3}{16} \rho^2 \left[ -\frac{g_1^8}{108 \alpha^2} \rho^4 \, dt^2 + \frac{8 g_1^2 \Xi_1}{3 \alpha} \sinh^2\!\theta \, dt dy + \sinh^2\!\theta \, dy^2 + d\theta^2 \right],
\feq
where we defined $\rho$ and $\theta$ by $r\equiv\frac{g_1^4\rho^4}{192\alpha}$,
$\coth x\equiv\cosh\theta$, and chose $\Xi_2=0$.

\section{Nonextremal rotating black holes in the $\overline{\mathbb{C}\text{P}}^n$ model}
\label{sec:non-extr-CPn}

In this section we shall construct a nonextremal deformation of the one quarter BPS solution presented in section~\ref{1/4BPS}. To this end we shall take a Carter-Pleba\'nski-type ansatz for the metric similar
to \eqref{CPPD}, where $Q(q)$ and $P(p)$ are quartic polynomials in $q$ and $p$ respectively,
\eq \label{nonextr-metric}
ds^2 = -\frac{Q}{W} \left(dt - p^2 dy\right)^2 + \frac{P}{W} \left(dt + (q^2-\Delta^2) dy\right)^2 + W\left( \frac{dq^2}{Q} + \frac{dp^2}{P} \right) \,,
\feq
\eq
Q = \sum_{n=0}^4 a_n q^n \,,  \qquad  P = \sum_{n=0}^4 b_n p^n \,,  \qquad  W = p^2 + q^2-\Delta^2 \,,
\feq
where $a_n$, $b_n$ and $\Delta$ are real constants. The ansatz for the scalars and the gauge potentials
is inspired by \eqref{CPscalars} and \eqref{CPAI}, 
\eq \label{nonextr-scalar}
z^\alpha = \frac1{p + i (q-\tilde\Delta)} \left(-\frac{g_\alpha}{g_0}(p+i q) + i c^\alpha\right)\,,
\feq
\eq \label{nonextr-vector}
A^I = \mathsf{P}^I \frac{p}{W} \left[dt + (q^2-\Delta^2) dy\right] \,,
\feq
with $\mathsf{P}^I$ real constants related to the magnetic charges, $\tilde\Delta$ real and $c^\alpha$ 
complex constants. Plugging these expressions into the equations of motion
\eqref{var-einstein}-\eqref{var-scalars} gives a set of constraints for the constants. It then turns out
that at a certain point one has to choose whether $\Delta$ vanishes or not. In what follows we shall
assume $\Delta\neq 0$, while the case $\Delta= 0$ is postponed to section~\ref{same-delta}.\\
For $\mathsf{P}^I$ not proportional to the coupling constants $g_I$ one class of solutions is obtained
by taking
\eq \label{nonextr-constants}
\begin{gathered}
a_0 = b_0 + b_2 \Delta^2 - 4 g^2 \Delta^4 - \frac{(g_I \mathsf{P}^I)^2}{2 g^2} + \frac{\mathsf{P}^2}{4} \,,
\qquad
a_1 = \frac{(g_I \mathsf{P}^I)\sqrt{(g_I \mathsf{P}^I)^2 - g^2 \mathsf{P}^2}}{2 g^2 \Delta}\,, \\
a_2 = -b_2 + 8g^2 \Delta^2 \,, \qquad  a_3 = 0\,, \qquad
a_4 = b_4 = -4g^2 \,,  \qquad  b_1 = b_3 = 0 \,, \\
\tilde\Delta  = \Delta  \frac{(g_I \mathsf{P}^I) g_0 + g^2 \mathsf{P}^0 }{ g_0 \sqrt{(g_I \mathsf{P}^I)^2 - g^2 \mathsf{P}^2}} \, , \qquad
c^\alpha = \Delta\frac{(g_I\mathsf{P}^I) g_\alpha - g^2 \mathsf{P}^\alpha}{g_0\sqrt{(g_I\mathsf{P}^I)^2
- g^2 \mathsf{P}^2}} \,.
\end{gathered}
\feq
Here we defined $\mathsf{P}^2 \equiv \eta_{IJ} \mathsf{P}^I \mathsf{P}^J$. Fixing the Fayet–Iliopoulos constants $g_I$ the solution depends on $n+4$ parameters $b_0$, $b_2$, $\Delta$ and $P^I$.
However, our ansatz is left invariant under the scale transformation
\eq\label{eq:scale-transf}
\begin{gathered}
p \rightarrow \lambda p \, , \qquad
q \rightarrow \lambda q \, , \qquad
t \rightarrow t/\lambda  \, , \qquad
y \rightarrow y/\lambda^3 \, ,  \\
\Delta \rightarrow \lambda \Delta \, , \qquad
a_n \rightarrow \lambda^{4-n} a_n \, , \qquad
b_n \rightarrow \lambda^{4-n} b_n \, ,
\end{gathered}
\feq
which reduces the number of independent parameters to $n+3$.\\
With a few lines of computation it is possible to show that this solution contains the one presented in \cite{Gnecchi:2013mja} for the prepotential $F = -i \tilde X^0 \tilde X^1$ (a tilde is introduced in order to distuinguish between the two solutions). In order to do so, we must consider the case of just one vector multiplet ($n=1$) and perform a symplectic rotation. In particular, introducing the symplectic vectors
\eq
\mathcal{G} = \left(\begin{array}{c} 0 \\ g_I \end{array}\right) \,,  \qquad
\mathcal{Q} = \left(\begin{array}{c} \mathsf{P}^I \\ 0 \end{array}\right) \,,
\feq
and the symplectic matrix
\eq \label{sympl-matrix}
T =
\left(\begin{array}{c|c}
\begin{matrix}
1 & 1 \\
1 & -1
\end{matrix} &
0 \\ \hline
0 &
\begin{matrix}
\frac 12 & \frac 12 \\
\frac 12 & -\frac 12
\end{matrix}
\end{array}\right) \,,
\feq
the solution for the rotated $F = -i \tilde X^0 \tilde X^1$ prepotential can be obtained from the same metric and gauge fields in~\eqref{nonextr-metric} and~\eqref{nonextr-vector}, but with the charges and gauge couplings replaced by their rotated counterparts according to $\mathcal{Q} = T \mathcal{\tilde Q}$ and $\mathcal{G} = T \mathcal{\tilde G}$. On the other hand, the scalar field is
\eq
\tilde\tau = \frac{\tilde X^0}{\tilde X^1} = \frac{1-z^1}{1+z^1} \,,
\feq
where $\tilde X^I$ belongs to the new symplectic section $\mathcal{\tilde V} = T^{-1} \mathcal{V}$.\\
In the supersymmetric, extremal limit we recover the solution presented in section~\ref{1/4BPS}. To this
end, the charge parameters need to be chosen proportional to the gauge couplings, $\mathsf{P}^I = \lambda \, \eta^{IJ} g_J$, with $\lambda \in \mathbb{R}$, hence the relations presented above simplify to
\eq
\begin{gathered}
a_0 = b_0 + b_2 \Delta^2 - 4 g^2 \Delta^4 - \frac{\lambda^2 g^2}{4} \,,  \qquad  a_1 = 0 \,, \qquad
a_2 = -b_2 + 8g^2 \Delta^2\,, \\
a_3 =0 \,, \qquad a_4 = b_4 = -4g^2 \,,  \qquad  b_1 = b_3 = 0 \,,
\end{gathered}
\feq
while on the other hand we are no more able to derive an explicit expression for $c^\alpha$, but we can
only assert that they must satisfy the conditions
\eq
-g_0 + c^\alpha g_\alpha = 0 \,, \qquad
c^\alpha c^\alpha = c^\alpha\bar{c}^\alpha = \tilde\Delta^2- \frac{\Delta^2 \, g^2}{ g_0^2} \,,
\feq
where summation over $\alpha$ is understood. Then, we see that the BPS solution \eqref{CPPD}, 
\eqref{CPscalars} and \eqref{CPAI} is recovered for
\eq \label{eq:extlimit}
L^2 = -\frac{1}{4g^2} \,,  \qquad  b_0 = (1+A) \frac{\mathsf{E}^2 L^2}{4} \,,  \qquad  b_2 = -\mathsf{E} \,,  \qquad  \lambda = 2 \, \mathsf{E} L^2 \sqrt{-A} \,,
\feq
and by identifying $\tilde\Delta=\Delta_1$ and $c^\alpha=-\Delta_1\beta^\alpha/\beta^0$.

\subsection{Properties of the compact horizon case}
\label{phys-disc}

Since $P$ is an even polynomial we may assume it has two distinct pairs of roots $\pm p_a$ and $\pm p_b$, where $0 < p_a < p_b$. We then consider solutions with $p$ in the range $|p| \le p_a$ by setting $p = p_a \cos\theta$, where $0 \le \theta \le \pi$, to obtain black holes with a compact horizon. We now use the scaling symmetry \eqref{eq:scale-transf} to set $p_b = L$ without loss of generality, where
$L^{-2} = -4g^2$. Defining the rotation parameter $j$ by $p_a^2 = j^2$, this means
\eq \label{b0b2}
b_0 = j^2\,,  \qquad  b_2 = -1 - \frac{j^2}{L^2} \, .
\feq
Then, after the coordinate transformation,
\eq \label{shift-t}
t \mapsto t + \frac{j\phi}{\Xi}\,,  \qquad  y \mapsto \frac{\phi}{j\Xi}\, ,
\feq
with $\Xi = 1 - \frac{j^2}{L^2}$, the metric~\eqref{nonextr-metric} becomes
\eq \label{nonextr-metric-spher}
\begin{aligned}
ds^2 = & -\frac{Q}{W} \left( dt + \frac{j\sin^2\!\theta}{\Xi} d\phi \right)^2 + \frac{\Delta_\theta \sin^2\!\theta}{W} \left( jdt + \frac{q^2-\Delta^2 + j^2}{\Xi}d\phi \right)^2 + \\
       & + W \left(\frac{dq^2}{Q} + \frac{d\theta^2}{\Delta_{\theta}}\right) \,,
\end{aligned}
\feq
where
\eqNN
W = q^2-\Delta^2 + j^2\cos^2\!\theta \,, \qquad
\Delta_\theta = 1 - \frac{j^2}{L^2}\cos^2\!\theta \,.
\feqNN
We notice that for zero rotation parameter, $j = 0$, \eqref{nonextr-metric-spher} boils down to the static nonextremal black holes with running scalar constructed in~\cite{Klemm:2012yg}, after the $n=1$
truncation and the symplectic rotation \eqref{sympl-matrix} are performed.\\
\eqref{nonextr-metric-spher} has an event horizon at $q=q_{\text h}$, where $q_{\text h}$ is the largest 
root of $Q$. The Bekenstein-Hawking entropy of the black hole is given by
\eq \label{entropy}
S = \frac{\pi}{\Xi G} \left(q^2_{\text h}-\Delta^2 + j^2\right) \,,
\feq
where $G$ denotes Newton's constant. In order to compute the temperature and angular velocity it is convenient to write the metric in the ADM form
\eq
ds^2 = -N^2 dt^2 + \sigma (d\phi - \omega dt)^2 + W \left(\frac{dq^2}{Q} + \frac{d\theta^2}{\Delta_\theta}\right) \,,
\feq
with
\eqNN
N^2 = \frac{Q\Delta_\theta W}{\Sigma^2} \,,  \qquad  \sigma = \frac{\Sigma^2\sin^2\!\theta}{W \Xi^2} \,,  \qquad  \omega = \frac{j\Xi}{\Sigma^2}[Q - \Delta_\theta (q^2-\Delta^2 + j^2)] \,,
\feqNN
where
\eqNN
\Sigma^2 = \Delta_\theta(q^2-\Delta^2 + j^2)^2 - Q j^2 \sin^2\!\theta \,.
\feqNN
The angular velocity at the horizon and at infinity are thus
\eq \label{omega}
\omega_{\text h} = -\frac{j\Xi}{q^2_{\text h}-\Delta^2 + j^2} \,,  \qquad  \omega_{\infty} = \frac{j}{L^2} \,.
\feq
The angular momentum may be computed as a Komar integral, which leads to
\eq \label{J}
J = \frac{a_1 j}{2\Xi^2 G}\,.
\feq
To get the mass of the solution we use the Ashtekar-Magnon-Das (AMD) formalism \cite{Ashtekar:1984zz,Ashtekar:1999jx}, applied to the conformally rescaled metric
$\bar{g}_{\mu\nu} = (L/q)^2 g_{\mu\nu}$. This gives
\eq \label{M}
M = -\frac{{a}_1}{2\Xi^2 G} \,.
\feq
Notice that the `ground state' $a_1 = 0$ represents a naked singularity. This can be seen as follows. The curvature singularity at $W = 0$\footnote{Note also that for $W < 0$, the real part of the scalar field becomes negative, so that ghost modes appear.} is shielded by a horizon if $q^2_{\text h} -\Delta^2 + j^2 \cos^2\!\theta > 0$, and thus $q^2_{\text h} > \Delta^2 $, which is equivalent to 
\eqNN
a_2^2 - \frac{4 a_0}{L^2} > \left(1 + \frac{j^2}{L^2}\right)^2 \,,
\feqNN
where we used the expression for $q_{\text h}$. This relation is easily shown to be violated for $a_1=0$
by using \eqref{nonextr-constants}.\\
The magnetic charges $\mathsf{p}^I$ are given by
\eq
\mathsf{p}^I = \frac{1}{4\pi} \oint_{\text{S}^2_{\infty}} F^I = -\frac{\mathsf{P}^I}{\Xi} \,.
\feq
The product of the horizon areas reads
\eq \label{area-prod}
\prod_{\Lambda=1}^4 A_\Lambda = \frac{(4\pi)^4}{\Xi^4} \prod_{\Lambda=1}^4 (q^2|_{\text h_\Lambda} -\Delta^2+ j^2) = (4\pi)^4 L^4 \left[ \frac{(\mathsf{p}^2)^2}{16} + 4 G^2 J^2 \right] \,,
\feq
where $\mathsf{p}^2 \equiv \eta_{IJ} \mathsf{p}^I \mathsf{p}^J$. In the second step we followed what has been done in~\cite{Gnecchi:2013mja} and the procedure explained in~\cite{Toldo:2012ec}. The charge-dependent term on the rhs of~\eqref{area-prod} is directly related to the prepotential; a fact that was first noticed in~\cite{Toldo:2012ec} for static black holes.\\
Now that we have computed the physical quantities of our solution, we see that one may choose the
$n+3$ free parameters as $\mathsf{P}^I, \Delta, j$, or alternatively $\mathsf{p}^I, M, J$. Our black holes are therefore labeled by the values of $n+1$ independent magnetic charges, the mass and the angular momentum.

\subsection{Thermodynamics and extremality}
\label{thermodynamics}

Imposing regularity of the Wick-rotated metric it is straightforward to compute the Hawking temperature,
with the result
\eq \label{temp}
T = \frac{Q_{\text h}'}{4\pi (q^2_{\text h} -\Delta^2 + j^2)} \,,
\feq
where $Q_{\text h}'$ denotes the derivative of $Q$ evaluated at the horizon.\\
Using the extensive quantities $S$, $M$, $J$ and $\mathsf{p}^I$ computed above,
it is possible to obtain the Christodoulou-Ruffini-type mass formula
\eq \label{CR}
\begin{aligned}
M^2 = \; & \frac{S}{4\pi G} + \frac{\pi J^2}{SG} + \frac{\pi}{4 S G^3} \frac{(\mathsf{p}^2)^2}{16} + \left(\frac{L^2}{G^2} + \frac{S}{\pi G}\right) \left((g_I \mathsf{p}^I)^2 + \frac{\mathsf{p}^2}{8 L^2}\right) + \\
         & + \frac{J^2}{L^2} + \frac{S^2}{2\pi^2 L^2} + \frac{S^3G}{4\pi^3 L^4} \,.
\end{aligned}
\feq
Since $S$, $ J$ and $\mathsf{p}^I$ form a complete set of extensive parameters, \eqref{CR} gives
the thermodynamic fundamental relation $M = M(S, J, \mathsf{p}^I)$. The intensive quantities
conjugate to $S$, $J$ and $\mathsf{p}^I$ are the temperature
\eq \label{T}
\begin{aligned}
T = \left.\frac{\partial M}{\partial S}\right|_{\!J,\mathsf{p}^I} = \;
    & \frac{1}{8\pi GM} \left[1 - \frac{4\pi^2 J^2}{S^2} - \frac{\pi^2}{S^2G^2} \frac{(\mathsf{p}^2)^2}{16} + 4\left((g_I \mathsf{p}^I)^2 + \frac{\mathsf{p}^2}{8 L^2}\right) + \right. \\
    & \left. + \frac{4SG}{\pi L^2} + \frac{3 S^2 G^2}{\pi^2 L^4} \right] \,,
\end{aligned}
\feq
the angular velocity
\eq \label{Omega}
\Omega = \left.\frac{\partial M}{\partial J}\right|_{\!S,\mathsf{p}^I} = \frac{\pi J}{MGS} \left[1 + \frac{SG}{\pi L^2} \right] \,,
\feq
and the magnetic potentials
\eq \label{magn-pot}
\begin{aligned}
\Phi_I = \left.\frac{\partial M}{\partial\mathsf{p}^I}\right|_{\!S,J,\mathsf{p}^{K \neq I}} = \;
         & \frac{1}{MG} \left[ \frac{\pi}{4S G^2} \frac{\mathsf{p}^2}{16} \, \eta_{IK} \mathsf{p}^K + \right. \\
         & \left. + \left(\frac{L^2}{G} + \frac{S}{\pi}\right) \left((g_K \mathsf{p}^K) g_I + \frac{1}{16 L^2} \, \eta_{IK} \mathsf{p}^K\right) \right] \,.
\end{aligned}
\feq
These quantities satisfy the first law of thermodynamics
\eq
dM = T \, dS + \Omega \, dJ + \Phi_I \, d\mathsf{p}^I \,.
\feq
It is straightforward to verify that expression~\eqref{T} for the temperature agrees with~\eqref{temp}, while from~\eqref{Omega} we observe that
\eq
\Omega = \omega_{\text h} - \omega_\infty \,,
\feq
with $\omega_{\text h}$ and $\omega_\infty$ given by~\eqref{omega}. Thus, what enters the first law
is the difference between the angular velocities at the horizon and at infinity.\\
Extremal black holes have vanishing Hawking temperature~\eqref{temp}, which happens when $q_{\text h}$ is at least a double root of $Q$. The structure function $Q$ can then be written as 
\eqNN
Q = (q-q_{\text h})^2 \left[ \frac{q^2}{L^2} +\frac{2 q_{\text h}}{L^2} q + a_2 + \frac{3q_{\text h}^2}{L^2} \right] \,,
\feqNN
so we must have
\eq\label{eq:extr-limit-a0-a1}
a_0 = a_2 q_{\text h}^2  + \frac{3q_{\text h}^4}{L^2} \,,  \qquad  a_1 = -2 a_2 q_{\text h} - \frac{4q_{\text h}^3}{L^2} \,.
\feq
It is straightforward to check that these relations are satisfied in the supersymmetric limit $P^I=\lambda \eta^{IJ}g_J$ previously described. On the other hand, it may happen that the free parameters are chosen such that \eqref{nonextr-constants} is compatible with \eqref{eq:extr-limit-a0-a1}, even if the charges are not proportional to the gauge couplings. In that case we would obtain an extremal, non-supersymmetric black hole.\\
To obtain the near-horizon geometry of the extremal black holes, we define new (dimensionless) coordinates $z, \hat t, \hat\phi$ by
\eq \label{zoom}
q = q_{\text h} + \epsilon q_0 z \,,  \qquad  t = \frac{q_0}{\Xi\epsilon}\hat t \,,  \qquad
\phi = \hat\phi + \frac{\omega_{\text h} q_0}{\Xi\epsilon}\hat t\,,
\feq
with
\eqNN
q_0^2 \equiv \frac{\Xi (q^2_{\text h}-\Delta^2 + j^2)}{C} \,,  \qquad  C = \frac{6 q_{\text h}^2}{L^2} + a_2 \,,
\feqNN
and take $\epsilon \to 0$ keeping $z, \hat t, \hat\phi$ fixed. This leads to
\eq
\begin{aligned}
ds^2 = & \; \frac{q^2_{\text h}-\Delta^2 + j^2 \cos^2\!\theta}{C} \left( -z^2 d\hat t^2 + \frac{dz^2}{z^2} + C \, \frac{d\theta^2}{\Delta_\theta} \right) + \\
       & + \frac{\Delta_\theta (q^2_{\text h}-\Delta^2 + j^2)^2 \sin^2\!\theta}{\Xi^2
(q^2_{\text h}-\Delta^2 + j^2 \cos^2\!\theta)} \left(d\hat\phi + \frac{2q_{\text h}\omega_{\text h}}C z \, d\hat t\right)^2\,,
\end{aligned}
\feq
where the constant $C$ is explicitly given by
\eqNN
C = \left[ \frac{(L^2 - \Delta^2)^2}{L^4} + \frac{(j^2 - \Delta^2)^2}{L^4} + 14 \, \frac{(L^2 - \Delta^2)(j^2 - \Delta^2)}{L^4} + 24(g_I \mathsf{P}^I)^2 + \frac{3 \mathsf{P}^2}{L^2} \right]^{1/2} \,.
\feqNN
Note that in the extremal limit it is manifest that the entropy is a function of the charges $J$ and $\mathsf{p}^I$ by solving \eqref{T} (for $T=0$) in terms of $S$.

\subsection{Case $\Delta= 0$}
\label{same-delta}

Solving the equations of motion with the Carter-Pleba\'nski-like ansatz~\eqref{nonextr-metric} and the assumption $\Delta=0$ leads to the relations
\eq \label{nonextr-constants-sameD}
\begin{gathered}
a_0 = b_0 - \frac{\mathsf{P}^2}{4} \,,  \qquad  a_2 = -b_2 \,, \qquad
a_3 = 0 \,,  \qquad  a_4 = b_4 = -4g^2 \,, \\
b_1 = b_3 = 0 \,, \qquad \tilde\Delta  =\frac{(g_I \mathsf{P}^I) g_0 + g^2 \mathsf{P}^0}{2 g_0 g^2 a_1} 
g_I \mathsf{P}^I\,, \qquad
c^\alpha = \frac{(g_I \mathsf{P}^I) g_\alpha - g^2 \mathsf{P}^\alpha}{2 g_0 g^2 a_1 } g_I \mathsf{P}^I\,.
\end{gathered}
\feq
Notice that in this case $a_1$ is not fixed by any condition, and remains thus a free parameter. Moreover
the equations of motion yield an additional condition on the charges, 
\eq\label{eq:cond-charges-Delta=0}
(g_I \mathsf{P}^I)^2 = g^2 \mathsf{P}^2\,.
\feq
This implies that the charges are proportional to the gauge couplings\footnote{To see this, choose in
$(n+1)$-dimensional Minkowski space with metric $\eta_{IJ}$ a basis in which the only nonvanishing
component of $g_I$ is $g_0$ (note that $g_I$ is timelike). Then \eqref{eq:cond-charges-Delta=0}
boils down to $\mathsf{P}^\alpha=0$.}. Nevertheless, notice that the solution is only supersymmetric if the free parameter $a_1$ is set to zero and the relations \eqref{eq:extlimit} hold. If $a_1 \neq 0$, the solution generalizes the Kerr-Newman-AdS black hole with $n$ magnetic charges and constant scalars. In order to 
shew this, one has to take $b_0$ and $b_2$ in the form \eqref{b0b2} and identify $a_1 = -2m$,
where $m$ and $j$ are the mass and angular momentum of the Kerr-Newman-AdS solution.\\
The mass, angular momentum and magnetic charges may be computed as in the case $\Delta\neq 0$,
which leads to the same expressions. The Christodoulou-Ruffini formula~\eqref{CR} is still valid, but with a simplification due to \eqref{eq:cond-charges-Delta=0},
\eq
M^2 = \frac{S}{4\pi G} + \frac{\pi J^2}{SG} + \frac{\pi}{4 S G^3} \frac{(\mathsf{p}^2)^2}{16} - \left(\frac{L^2}{G^2} + \frac{S}{\pi G}\right) \frac{\mathsf{p}^2}{8 L^2} + \frac{J^2}{L^2} + \frac{S^2}{2\pi^2 L^2} + \frac{S^3G}{4\pi^3 L^4} \,.
\feq
This relation reduces correctly to equation~(43) of~\cite{Caldarelli:1999xj} in the KNAdS case if we
identify $\mathsf{p}^2 = -4Q^2$.


\section*{Acknowledgements}

This work was supported partly by INFN.

\appendix

\section{Equations of motion}
The equations of motion following from \eqref{action} are given by
\eq \label{var-einstein}
R_{\mu\nu} = -(\text{Im}\,{\cal N})_{IJ} F^I_{\mu\lambda} F^{J\ \lambda}_{\ \nu} + \frac{1}{4} g_{\mu\nu} (\text{Im}\,{\cal N})_{IJ} F^I_{\rho\sigma} F^{J\rho\sigma} + 2 g_{\alpha\bar\beta} \partial_{(\mu}z^{\alpha}\partial_{\nu)}\bar z^{\bar\beta} + g_{\mu\nu} V \,,
\feq
\eq \label{var-maxwell}
\nabla_\mu\left[(\text{Im}\,{\cal N})_{IJ} F^{J\mu\nu} - \frac{1}{2} (\text{Re}\,{\cal N})_{IJ} \, e^{-1} \epsilon^{\mu\nu\rho\sigma} F^J_{\rho\sigma}\right] = 0 \,,
\feq
\eq \label{var-scalars}
\begin{split}
\frac{1}{4} \frac{\delta(\text{Im}\,{\cal N})_{IJ}}{\delta z^\alpha} & F^I_{\mu\nu} F^{J\mu\nu} - \frac{1}{8} \frac{\delta(\text{Re}\,{\cal N})_{IJ}}{\delta z^\alpha} \, e^{-1} \epsilon^{\mu\nu\rho\sigma} F^I_{\mu\nu} F^J_{\rho\sigma} + \frac{\delta g_{\alpha\bar\beta}}{\delta \bar z^{\bar\gamma}} \partial_\lambda \bar z^{\bar\gamma} \partial^\lambda \bar z^{\bar\beta} \\
& + g_{\alpha\bar\beta} \nabla_\lambda \nabla^\lambda \bar z^{\bar\beta} - \frac{\delta V}{\delta z^\alpha} = 0 \,,
\end{split}
\feq
which hold for any prepotential $F$.

\end{document}